\documentclass[10pt,twocolumn,preprintnumbers,amsmath,amssymb,nofootinbib
,superscriptaddress]{revtex4}
\usepackage{graphicx,longtable,mathrsfs,color,array}
\usepackage[hidelinks]{hyperref}
\usepackage[usenames,dvipsnames]{xcolor} 
\usepackage{amssymb,amsmath,mathtools,mathrsfs,slashed} 
\usepackage{epsfig,subfigure,placeins,float} 
\usepackage{booktabs,longtable,ctable,multirow} 
\usepackage{exscale,relsize} 
\usepackage[normalem]{ulem} 
\usepackage{enumerate}
\usepackage{times, mathptmx} 
\usepackage[utf8]{inputenc}
\usepackage{color}
\usepackage{hyperref}
\usepackage{graphicx}
\usepackage{color}
\usepackage{graphicx,graphics}
\allowdisplaybreaks[1]

\newcommand{\be}{\begin{equation}}
\newcommand{\ee}{\end{equation}}
\newcommand{\bea}{\begin{eqnarray}}
\newcommand{\eea}{\end{eqnarray}}

\usepackage[stable]{footmisc}
\usepackage{color}

\def\dkmu2{\delta K_{\mu \nu}\delta K^{\mu \nu}}
\def\pmu2{ \phi_{\mu \nu}\phi^{\mu \nu}}

\begin{document}

\title{A covariant approach to parameterised cosmological perturbations}
\author{Oliver J. Tattersall}
\email{oliver.tattersall@physics.ox.ac.uk}
\affiliation{Astrophysics, University of Oxford, DWB, Keble Road, Oxford OX1 3RH, UK}
\author{Macarena Lagos}
\affiliation{Astrophysics, University of Oxford, DWB, Keble Road, Oxford OX1 3RH, UK}
\affiliation{Theoretical Physics, Blackett Laboratory, Imperial College London, Prince Consort Road, London SW7 2BZ, UK}
\author{Pedro G. Ferreira}
\affiliation{Astrophysics, University of Oxford, DWB, Keble Road, Oxford OX1 3RH, UK}
\date{Received 30 June 2017; published 8 September 2017}

\begin{abstract}
We present a covariant formulation for constructing general quadratic actions for cosmological perturbations, invariant under a given set of gauge symmetries for a given field content. This approach allows us to analyse scalar, vector and tensor perturbations at the same time in a straightforward manner. We apply the procedure to diffeomorphism invariant single-tensor, scalar-tensor and vector-tensor theories and show explicitly the full covariant form of the quadratic actions in such cases, in addition to the actions determining the evolution of vector and tensor perturbations. We also discuss the role of the symmetry of the background in identifying the set of cosmologically relevant free parameters describing these classes of theories, including calculating the relevant free parameters for an axisymmetric Bianchi-I vacuum universe.
\end{abstract}


\maketitle
\section{Introduction}
Einstein's theory of General Relativity (GR) has survived numerous tests throughout over 100 years of its existence \cite{Will:2014kxa}, including the recent discovery of gravitational waves from the merger of a black hole binary system \cite{Abbott:2016blz}. There are still, however, compelling arguments for proposing and testing alternative models to, and extensions of, Einstein's GR \cite{Berti:2015itd, Clifton20121}. The prospect of constraining GR with future surveys (\cite{Amendola:2012ys} for example) provides motivation for developing a method for parameterising deviations from GR in as general a way as possible. This has been done in the weak field regime successfully through the Parameterised Post Newtonian (PPN) formalism \cite{Berti:2015itd,Sanghai:2016tbi}, whilst the Parameterised Post Friedmannian formalism \cite{Baker:2011jy,Baker:2012zs} and other approaches \cite{Gubitosi:2012hu,Bloomfield:2012ff,Creminelli:2008wc,Gleyzes:2013ooa,Gleyzes:2014rba,Gleyzes:2014dya,Battye:2012eu} have been proposed to test GR on cosmological scales. 

In \cite{Lagos:2016wyv,2017JCAP...01..047L} a method was proposed for constructing a general action, quadratic in the perturbed gravitational fields around a cosmological background, for general theories of gravity. Such an action depends on a finite number of free, time dependent, functions. It was shown that the presence of gauge symmetries leads to a number of Noether constraints which greatly restricts the number of free functions in such a way that the resulting action describes the most general gauge invariant action which is quadratic in the perturbed gravitational fields.

The method of \cite{Lagos:2016wyv,2017JCAP...01..047L} was applied to an expanding Friedmann-Lemaitre-Robertson-Walker (FLRW) universe, using the Arnowitt-Deser-Misner (ADM) variables as building blocks for the action. In this paper we want to reach out and consider how one might construct a formalism which could be applied easily to a general background. The way is to consider a \textit{fully covariant} method where we again use the power of the Noether constraints to impose the presence of gauge symmetries in the theory. Furthermore, by constructing actions which are fully covariant one can consider scalar, vector, and tensor type perturbations all in one go, as opposed to solely scalar perturbations, as in \cite{Lagos:2016wyv,2017JCAP...01..047L}. In this paper we will, as in \cite{Lagos:2016wyv,2017JCAP...01..047L} focus on a homogeneous and isotropic, cosmological background. But throughout, we will discuss the main lessons which will allow us to consider more general background space-times. A key aspect of this paper is that we will discuss the role that the symmetries of the background have and their interplay with gauge invariance. In doing so, we prepare the ground for more general analyses of linear perturbations in relativistic theories of gravity.

\textit{Outline:} In Section \ref{SecFP} we recap the method, now in the covariant form and use it to derive the action of a free massless spin-2 field propagating on Minkowski space, which corresponds to linearised GR. In Sections \ref{secGR}-\ref{secVT}, we will derive the diffeomorphism-invariant quadratic actions of linear perturbations on a FLRW background for three families of theories of gravity: containing a single tensor field, a tensor field with a scalar field, and a tensor field with a vector field, respectively. The results of this paper can thus be compared to those of \cite{Lagos:2016wyv} and \cite{Gleyzes:2014rba}. In Section \ref{Symmetries} we discuss how the symmetries of the background impact the number of free functions characterising the resulting gravitational action. In particular, we will repeat the calculation of Section \ref{secGR} with an axisymmetric Bianchi-I background \cite{Ellis:1968vb}. In Section \ref{conclusion} we will discuss the results of our work and the method presented in this paper, as well as future work to be undertaken. 

Throughout this paper, indices using the greek alphabet ($\mu$, $\nu$, $\lambda$...) will denote space-time indices and run over coordinates 0-3. Roman letters ($i$, $j$, $k$...) will denote spatial indices and run over coordinates 1-3. The metric signature will be $(-,+,+,+)$.


\section{Covariant action approach}\label{SecFP}
In this section we describe the covariant method for constructing gauge invariant quadratic actions for linear perturbations and illustrate it by recovering linear general relativity in Minkowski space. We discuss the role of the global symmetry of the background and the local gauge symmetry of the perturbations in the method. 

We follow the same logic as in \cite{Lagos:2016wyv,2017JCAP...01..047L} but using a covariant approach. The main steps of the method are summarised as follows: 
\begin{enumerate}
\item For a given set of gravitational fields, choose a background and write a set of covariant projectors (a set of vectors and tensors) that foliate your space-time following the global symmetries of the background. Then, consider linear perturbations for each gravitational (and matter) field.

\item Construct the most general quadratic action for the gravitational fields by writing all possible compatible contractions of the covariant background projectors and the linear perturbations. Introduce a free function of the background in front of each possible term and truncate the number of possible terms in the action by choosing a maximum number of derivatives.

\item Choose a desired gauge symmetry and impose local invariance of the quadratic action by solving a set of Noether constraints. The resulting action will be the most general quadratic gauge invariant action around a background with a given set of global symmetries. 

\end{enumerate} 

We now proceed to illustrate the method by following each one of the previous step in the case of a single tensor gravitational field $g_{\mu\nu}$ (or metric) in vacuum with a diffeomorphism invariant action. In this case, the background will correspond to Minkowski space: 
\begin{equation}
\bar{g}_{\mu\nu}=\eta_{\mu\nu},
\end{equation}
where the bar denotes the background value of the metric, and $\eta_{\mu\nu}=\text{diag}(-1,1,1,1)$ is the Minkowski metric. We know that this background has a global symmetry under the Poincare group, and thus we can describe the metric with only one projector, the tensor $\eta_{\mu\nu}$, that follows this symmetry. Hence, in this case, we do not need to make any particular foliation. Next, we consider linear perturbations and thus the full metric can be expressed as:
\begin{equation}
g_{\mu\nu}=\eta_{\mu\nu}+ h_{\mu\nu}; \quad | h_{\mu\nu}| \ll |\eta_{\mu\nu}|,
\end{equation}
where $h_{\mu\nu}$ is a linear perturbation, which can be a function of space and time.

We now follow step 2 and write the most general covariant quadratic action leading to second-order derivative equations of motion. In this case, we can only have two different possible terms (modulo total derivatives):
\begin{equation}\label{GRgral}
S^{(2)}=\int d^4x \left[ \mathcal{A}^{\mu\alpha\beta\nu\gamma\delta}\bar{\nabla}_\mu h_{\alpha\beta} \bar{\nabla}_\nu h_{\gamma\delta} + \mathcal{B}^{\alpha\beta\gamma\delta}h_{\alpha\beta}h_{\gamma\delta}\right],
\end{equation} 
where $\bar{\nabla}_\mu$ are covariant derivatives with respect to the background metric, and the coefficients $\mathcal{A}$ and $\mathcal{B}$ are arbitrary tensors, functions of the background. These tensors must respect the symmetries of the background and hence be constructed with the tensor $\eta_{\mu\nu}$. Explicitly, the most general form these tensors can take is the following:
\begin{align}\label{GRCoeffs}
\mathcal{A}^{\mu\alpha\beta\nu\gamma\delta} = & c_3 \eta^{\mu\nu}\eta^{\alpha\beta}\eta^{\gamma\delta}+ c_4 \eta^{\mu\alpha}\eta^{\nu\beta}\eta^{\gamma\delta} \nonumber \\ &+ c_5\eta^{\mu \nu}\eta^{\alpha\gamma}\eta^{\beta\delta} + c_6\eta^{\mu\gamma}\eta^{\nu\alpha}\eta^{\beta\delta},  \nonumber\\
\mathcal{B}^{\alpha\beta\gamma\delta} = & c_1 \eta^{\alpha\beta}\eta^{\gamma\delta}+ c_2 \eta^{\alpha\gamma}\eta^{\beta\delta},
\end{align}
where the coefficients $c_n$ are free functions of the background, i.e.~constants in this case. We note that we have not actually written all the possible contractions in these tensors $\mathcal{A}$ and $\mathcal{B}$, but instead only those that are inequivalent after considering the contraction with the symmetric tensor perturbation $h_{\mu\nu}$ in the action in eq.~(\ref{GRgral}).


If we separate each term of the action explicitly, the resulting most general quadratic action takes the following form:
\begin{align}
  S^{(2)}= \int d^4x&\left[ c_1h^2+c_2h_{\mu\nu}h^{\mu\nu}+c_3\partial_\mu h \partial^\mu h +c_4 \partial_\mu h^{\mu\nu}\partial_\nu h \right.\nonumber \\ & \ \ \ \left.+c_5\partial_\mu h_{\nu\lambda}\partial^\mu h^{\nu\lambda}+c_6\partial_\mu h_{\nu\lambda}\partial^\nu h^{\mu\lambda}\right], \label{SFP}
\end{align}
where $h=\eta^{\mu\nu}h_{\mu\nu}$ and indices are lowered and raised with the background metric $\eta_{\mu\nu}$.

We now proceed to follow step 3, and we will impose symmetry under linear diffeomorphism invariance. Consider an infinitesimal coordinate transformation:
\begin{align}
  x^{\mu}\rightarrow x^\mu + \epsilon^\mu;\quad  |\epsilon^\mu | \ll | x^\mu|, \label{coordtransformation}
\end{align}
where $\epsilon^\mu$ is a linear perturbation that depends on space and time. Under this transformation the background stays the same but the gravitational perturbation field changes as:
\begin{align}
  h_{\mu\nu}\rightarrow h_{\mu\nu}+\partial_\mu\epsilon_\nu+\partial_\nu\epsilon_\mu. \label{htransformation}
\end{align}
If we wish our theory to be invariant under this coordinate transformations, then the variation of the action in eq.~(\ref{SFP}) with respect to the transformation in eq.~(\ref{htransformation}) should vanish. After making suitable integrations by parts, we find that the variation of the action gives:
\begin{align}
  \delta_\epsilon S^{(2)}=\int &d^4x\,\epsilon_\mu \left[-4c_2\partial_\nu h^{\mu\nu}-4c_1\partial^\mu h\right. \nonumber \\&+ 2\, (c_4+c_6)\partial^\mu\partial^\nu\partial^\lambda h_{\nu\lambda} +2\, (2c_5+c_6)\partial_\nu\Box h^{\mu\nu} \nonumber\\
  &\left.+2\, (2c_3+c_4)\partial^\mu\Box h\right], \label{deltaSFP}
\end{align}
where $\Box=\eta^{\mu\nu}\partial_\mu\partial_\nu$ is the d'Alembertian operator. For the action to be gauge invariant we need $\delta_\epsilon S^{(2)}$ to vanish for arbitrary $\epsilon^\mu$, and therefore the whole integrand to vanish. This leads to the following Noether identity:
\begin{eqnarray}
-4c_2\partial_\nu h^{\mu\nu}-4c_1\partial^\mu h+2\, (c_4+c_6)\partial^\mu\partial^\nu\partial^\lambda h_{\nu\lambda} &\nonumber \\ +2\, (2c_5+c_6)\partial_\nu\Box h^{\mu\nu} +2\, (2c_3+c_4)\partial^\mu\Box h &=0.
\end{eqnarray}
Since this must be satisfied off-shell, terms with different derivative structure must vanish independently, leading to the following set of Noether constraints:
\begin{align}
  c_1= & c_2=0,\nonumber\\
  c_4= & -c_6=-2c_3=2c_5.
\end{align}
These constraints are simple algebraic relations on the free coefficients $c_n$, and they ensure the action (\ref{SFP}) is diffeomorphism invariant. Using our freedom to rescale the size of $h_{\mu\nu}$, we can set $-4c_4=M_{Pl}^2$, the reduced Planck mass (squared), and write the resulting quadratic action as:
\begin{align}
  S^{(2)}= \int d^4x\,\frac{M_{Pl}^2}{4}&\left[ \frac{1}{2}\partial_\mu h \partial^\mu h - \partial_\mu h^{\mu\nu}\partial_\nu h \right. \nonumber \\ & \ \ \left.-\frac{1}{2}\partial_\mu h_{\nu\lambda}\partial^\mu h^{\nu\lambda}+\partial_\mu h_{\nu\lambda}\partial^\nu h^{\mu\lambda}\right], \label{SFPfinal}
\end{align}
which we recognise as the quadratic expansion of the Einstein-Hilbert action about a Minkowski background \cite{2011PhRvD..83f4038S}.



\section{Recovering general relativity in an expanding background}\label{secGR}
In this section we apply the covariant method for constructing a quadratic action for linear perturbations around a homogeneous and isotropic background, in the case where the gravitational field content is given by a single tensor field and the action is invariant under linear coordinate transformations. In this case, we will need to explicitly couple a matter sector to the gravitational action in order to have a non-trivial background solution. For simplicity and concreteness, let us consider a scalar field $\varphi$ minimally coupled to the gravitational tensor field, although the structure of the final gravitational quadratic action will be valid for a general perfect fluid as well. We emphasise that the procedure will give us a general parametrised gauge invariant gravitational action for cosmological perturbations, whereas the matter action is assumed to be known.

We start by following step 1. We assume that the background is given by a spatially flat FRW metric:
\begin{equation}
\bar{g}_{\mu\nu}=-dt^2+a(t)^2\delta_{ij}dx^idx^j,
\end{equation}
where $a(t)$ is the scale factor as a function of the physical time $t$. In order to describe this background in a covariant way, we make a 1+3 split and foliate the space-time with a time-like unit vector $u^\mu$, which induces orthogonal hypersurfaces with a spatial metric $\gamma_{\mu\nu}$ such that:
\begin{align}
\gamma_{\mu\nu}=\bar{g}_{\mu\nu}+u_\mu u_\nu. \label{gammadef}
\end{align}
Thus $u^\mu$ and $\gamma_{\mu\nu}$ act as the projectors for this space-time. Specifically in this case, the time-like vector and spatial metric are given by:
\begin{align}
u_\mu = & (-1,\bf{0})_\mu,\\
\gamma_{ij} = & a^2\delta_{ij},\\
\gamma_{\mu0} = & 0,
\end{align}
such that $\gamma_{\mu\nu}$ and $u_\mu$ are orthogonal to one another:
\begin{align}
  \gamma^{\mu\nu}u_\nu= 0.
\end{align}
A covariant 1+3 approach to cosmology has previously been developed \cite{Tsagas:2007yx} which shares similar projectors to the $u$ and $\gamma$ presented here. We, however, believe that strength of the formalism used in this paper is that it can be readily utilised to split the background spacetime in different ways  (e.g. see section \ref{Symmetries} for a 1+1+2 split). Further differences between the two formalisms are discussed in appendix \ref{appendixS1}.

We now add a matter scalar field $\varphi$, whose background must also be homogeneous and isotropic and hence can only be a function of time:
\begin{equation}
\bar{\varphi}= \bar{\varphi}(t).
\end{equation}
Next, we consider linear perturbations in the metric and matter field so the full perturbed fields are given by:
\begin{align}\label{TensorMatterLinearPerts}
g_{\mu\nu}=& \bar{g}_{\mu\nu}+h_{\mu\nu}; \quad |h_{\mu\nu}|\ll |\bar{g}_{\mu\nu}|,\nonumber\\
\varphi=& \bar{\varphi}(t)+\delta \varphi;\quad |\delta \varphi|\ll |\bar{\varphi}|.
\end{align}

We now move onto step 2 and construct the most general quadratic gravitational action. As in Section \ref{SecFP}, the most general action quadratic in $h_{\mu\nu}$ with up to second order equations of motion can be written as:
\begin{align}
S_G^{(2)}=\int d^4x\,a^3\;&\left[ \mathcal{A}^{\mu\nu\alpha\beta} h_{\mu\nu} h_{\alpha\beta}
 + \mathcal{B}^{\mu\nu\alpha\beta\delta}\bar{\nabla}_{\delta} h_{\mu\nu} h_{\alpha\beta}
\right. \nonumber \\ &\left. + \mathcal{C}^{\mu\nu\alpha\beta\kappa\delta}\bar{\nabla}_{\kappa} h_{\mu\nu} \bar{\nabla}_{\delta} h_{\alpha\beta}\right], \label{Sgencosmology}
\end{align}
where the coefficients $\mathcal{A}$, $\mathcal{B}$, and $\mathcal{C}$ are tensors depending on the background. Notice that here we have added a tensor with five indices $\mathcal{B}^{\mu\nu\alpha\beta\delta}$, which we ignored in the previous section as in Minkowski space this tensor would be constant and hence the second term in eq.~(\ref{Sgencosmology}) would correspond to a boundary term. Also, for future convenience we have defined the tensors in action (\ref{Sgencosmology}) with a factor $a^3$ in front. 

We now write the most general form that the tensors $\mathcal{A}$, $\mathcal{B}$, and $\mathcal{C}$ can have respecting the symmetries of the background. In this case, they can be constructed using the projectors, the time-like unit vector $u^\mu$ and the spatial metric $\gamma_{\mu\nu}$, in the following way:
\begin{align}
\mathcal{A}^{\mu\nu\alpha\beta}=\,&A_1\gamma^{\mu\nu}\gamma^{\alpha\beta}+ A_2\gamma^{\mu\alpha}\gamma^{\nu\beta}+ A_3\gamma^{\mu\nu}u^\alpha u^\beta \nonumber \\ & + A_4 \gamma^{\mu\alpha}u^\nu u^\beta+ A_5u^\mu u^\nu u^\alpha u^\beta, \label{Atensorcosmology}\\\medskip
\mathcal{B}^{\mu\nu\alpha\beta\delta}=\,& B_1\gamma^{\delta\alpha}\gamma^{\nu\beta}u^\mu + B_2\gamma^{\alpha\beta}\gamma^{\nu\delta}u^\mu + B_3\gamma^{\mu\nu}u^\alpha u^\beta u^\delta\nonumber\\
&+ B_4\gamma^{\mu\delta}u^\nu u^\alpha u^\beta, \label{Btensorcosmology}\\ \medskip
\mathcal{C}^{\mu\nu\alpha\beta\kappa\delta}=\,&C_1\gamma^{\mu\nu}\gamma^{\alpha\beta}\gamma^{\kappa\delta}+ C_2\gamma^{\mu\alpha}\gamma^{\nu\beta}\gamma^{\kappa\delta}+ C_3\gamma^{\mu\nu}\gamma^{\alpha\kappa}\gamma^{\beta\delta}\nonumber \\ &+ C_4\gamma^{\mu\kappa}\gamma^{\alpha\beta}\gamma^{\nu\delta}\nonumber\\
& + (C_5 \gamma^{\mu\nu}\gamma^{\alpha\beta}+ C_6\gamma^{\mu\alpha}\gamma^{\nu\beta})u^\kappa u^\delta \nonumber \\ &+  (C_7 \gamma^{\mu\nu}\gamma^{\kappa\delta}+ C_8\gamma^{\mu\kappa}\gamma^{\nu\delta})u^\alpha u^\beta  \nonumber\\
&+ C_{9}\gamma^{\alpha\beta} u^\mu u^\nu u^\kappa u^\delta+C_{10}\gamma^{\kappa\delta} u^\alpha u^\beta u^\mu u^\nu\nonumber\\
 &+ (C_{11}\gamma^{\kappa\delta}\gamma^{\beta \nu}+ C_{12}\gamma^{\kappa\beta}\gamma^{\delta \nu})u^\mu u^\alpha \nonumber \\ &+ (C_{13}\gamma^{\alpha\beta}\gamma^{\nu \delta}+ C_{14}\gamma^{\alpha\nu}\gamma^{\delta \beta})u^\mu u^\kappa\nonumber\\
 &+C_{15}\gamma^{\mu\alpha}u^\nu u^\beta u^\kappa u^\delta  + C_{16}\gamma^{\mu\kappa}u^\nu u^\beta u^\alpha u^\delta
 \nonumber\\
 &+ C_{17} u^\mu u^\alpha u^\nu u^\beta u^\kappa u^\delta, \label{Ctensorcosmology}
\end{align}
where, as in the previous section, we have only defined the set of tensors that lead to distinct terms in the quadratic action\footnote{Whilst in principle one should symmetrise over the indices of $\mathcal{A}$, $\mathcal{B}$, and $\mathcal{C}$ in order to obtain the most general tensors, the additional symmetrised terms do not contribute any new terms to the action so they have been ommited.}. Here, the coefficients $A_i$, $B_i$, and $C_i$ are arbitrary scalar functions of the background, and hence of time. We note that the tensors $\mathcal{A}$, $\mathcal{B}$, and $\mathcal{C}$ could come from the background metric $\bar{g}_{\mu\nu}$ and its derivatives to arbitrary order. Hence, we are restricting the number of derivatives allowed for the perturbations $h_{\mu\nu}$, but not for the background. 

From equations (\ref{Atensorcosmology})-(\ref{Ctensorcosmology}) we can see how less symmetric backgrounds can lead to a larger number of free parameters in the gravitational action. Whereas in Minkowski the action in step 2 had only 6 free constant parameters, in a homogeneous and isotropic background we find 26 free functions of time. As we shall see later, we will also find more Noether constraints in this section, and so the total gauge invariant action will have only one extra free parameter compared to the Minkowski case.

Having obtained an explicit expression for the coefficients in eq.~(\ref{Sgencosmology}), we proceed to step 3. We want the total action (gravity and matter) to be linearly diffeomorphism invariant. Specifically, we will impose gauge invariance in the gravitational action (\ref{Sgencosmology}) coupled to the matter action of a minimally coupled scalar field without a potential:
\begin{equation}
S_M=- \int d^4x \sqrt{-g}\left(\frac{1}{2}\bar{\nabla}_\mu \varphi\bar{\nabla}^\mu\varphi\right).
\end{equation} 
If we expand this action to quadratic order in the linear perturbations given in eq.~(\ref{TensorMatterLinearPerts}) we get:
\begin{align}
S_M^{(2)} = &-\int d^4x\,a^3 \left[\frac{1}{4}\left(\frac{1}{2}h^2-h_{\mu\nu}h^{\mu\nu}\right)\left(\frac{1}{2}\bar{g}^{\mu\nu}\bar{\nabla}_\mu \bar{\varphi}\bar{\nabla}_\nu\bar{\varphi}\right) \right.\nonumber \\ &\left. + \frac{1}{2}h\left(-\frac{1}{2}h^{\mu\nu}\bar{\nabla}_\mu \bar{\varphi}\bar{\nabla}_\nu\bar{\varphi} + \bar{g}^{\mu\nu}\bar{\nabla}_\mu \delta\varphi\bar{\nabla}_\nu\bar{\varphi} \right) \right. \nonumber \\ &\left.+ \left(-h^{\mu\nu}\bar{\nabla}_\mu\delta\varphi\bar{\nabla}_\nu\bar{\varphi} + \frac{1}{2}\bar{g}^{\mu\nu}\bar{\nabla}_\mu\delta\varphi\bar{\nabla}_\nu\delta\varphi \right.\right. \nonumber \\ & \ \ \ \ \left.\left.+ \frac{1}{2}hh^{\mu\nu}\bar{\nabla}_\mu\bar{\varphi}\bar{\nabla}_\nu\bar{\varphi}\right)\right], \label{Smattercosmology}
\end{align}
where we have defined the trace $h=h_{\mu\nu}\bar{g}^{\mu\nu}$. The matter action also leads to the following background equation of motion:
\begin{align}
\bar{\nabla}_\mu\bar{\nabla}^\mu\bar{\varphi}=0. \label{phiEOM}
\end{align}

We want the total quadratic action to be linearly diffeomorphism invariant. In this case, the metric perturbation will transform as the Lie derivative of the background metric along an infinitesimal coordinate transformation vector $\epsilon^\mu$. That is,
\begin{align}
 h_{\mu\nu}\rightarrow h_{\mu\nu}+\bar{\nabla}_\mu\epsilon_\nu+\bar{\nabla}_\nu\epsilon_\mu. \label{hgaugetransformation}
\end{align}
In addition, the matter scalar perturbation will transform as:
\begin{equation}\label{mattertransformation}
  \delta\varphi\, \rightarrow \, \delta\varphi+\epsilon^\mu\bar{\nabla}_\mu\bar{\varphi}.
\end{equation}

The total action given by the combination of (\ref{Sgencosmology}) and (\ref{Smattercosmology}) can now be varied to find the Noether identities. Schematically, an infinitesimal variation of the total action can be written as:
\begin{equation}
\hat{\delta} S^{(2)}_T=\hat{\delta} S^{(2)}_G+ \hat{\delta} S^{(2)}_M=\int d^4x \left[ \mathcal{E}^{\mu\nu} \hat{\delta} h_{\mu\nu}+ \mathcal{E}_{\varphi}\hat{\delta} (\delta \varphi)  \right],
\end{equation}
where $\hat{\delta}$ denotes a function variation, and $\mathcal{E}^{\mu\nu}$ with $\mathcal{E}_{\varphi}$ denote the equations of motion of the perturbation fields $h_{\mu\nu}$ and $\delta \varphi$, respectively. We now consider the functional variation of the action when the perturbation fields transform under the gauge symmetry, as in equations (\ref{hgaugetransformation}) and (\ref{mattertransformation}). After making suitable integrations by parts we find:
\begin{equation}\label{GRActionVariation}
\hat{\delta}_\epsilon S^{(2)}_T= \int d^4x\left[ -2\bar{\nabla}_\nu\left(\mathcal{E}^{\mu\nu}\right)+ \mathcal{E}_{\varphi} \bar{\nabla}^\mu \bar{\varphi}\right]\epsilon_\mu,
\end{equation}
where we have used the fact that $\mathcal{E}^{\mu\nu}$ is a symmetric tensor. For the total action to be gauge invariant we impose $\hat{\delta}_\epsilon S^{(2)}_T=0$, which leads to four Noether identities given by each one of the components of the bracket in eq.~(\ref{GRActionVariation}). From these Noether identities we can read a number of Noether constraints that will relate the values of the free parameters $A_i$, $B_i$ and $C_i$ of the quadratic gravitational action. In order to read off the Noether constraints easily, we rewrite the Noether identities solely in terms of the projectors $u^\mu$ and $\gamma_{\mu\nu}$, by eliminating all covariant derivatives of the background using the equations in Appendix \ref{appendixS1}. For instance, we will rewrite the covariant derivative of the background matter field as:
\begin{align}
\bar{\nabla}_\mu\bar{\varphi}=-u_\mu\dot{\bar{\varphi}},
\end{align}
where an overdot represents a derivative with respect to the physical time. In this way, due to the fact that $\gamma_{\mu\nu}$ and $u^\mu$ are orthogonal, any perturbation field contracted with tensors having different $u$ index structure, for example, must vanish independently. Through this process, the following Noether constraints are obtained for the $A_i$, $B_i$, and $C_i$:
\begin{align}
A_1&=-\frac{\dot{\bar{\varphi}}^2}{16},\nonumber\\
A_2&=\frac{1}{8}(\dot{\bar{\varphi}}^2+32H^2C_5+32H\dot{C}_5),\nonumber\\
A_3&=\frac{1}{8}(-\dot{\bar{\varphi}}^2+32H\dot{C}_5),\nonumber\\
A_4&=\frac{1}{4}(\dot{\bar{\varphi}}^2+(32H^2-16\dot{H})C_5+32H^2\dot{C}_5),\nonumber\\
A_5&=\frac{1}{16}(-\dot{\bar{\varphi}}^2-96H^2C_5),\nonumber\\
B_1&=4HC_5+4\dot{C}_5, \nonumber\\
B_3&=-B_4=4HC_5, \nonumber\\
C_1&=-C_2=-(C_5+H^{-1}\dot{C}_5), \nonumber\\
C_3&=-C_4=2C_5-2H^{-1}\dot{C}_5, \nonumber\\
2C_6&=C_8=-C_7=C_{11}=-C_{12}=-2C_5, \nonumber\\
C_{13}&=-C_{14}=-4C_5, \nonumber\\
B_2&=C_{15}=C_{16}=C_{17}=C_9=C_{10}=0.\label{cosNC}
\end{align}
In addition, we also obtain a constraint on the background quantities: 
\begin{align}
\dot{H}&=\frac{\dot{\bar{\varphi}}^2}{16C_5} \label{fried1}.
\end{align}
We note that this equation has a similar form to the Friedmann equation if we identify $-8C_5=M^2$, with the exception that $M$ is a free function of time here, as opposed to the constant Planck mass. We find that the number of free parameters in the action is reduced from 26 to only one. We note that the background also has one free function $a$ ($\bar{\varphi}$ is not considered free as it is related to $a$ through (\ref{phiEOM})), and that $a$ and $M$ are related through eq.~(\ref{fried1}). Hence the entire model (background and perturbations) is described using one free function. Explicitly, we find the total quadratic action to be given by:
\begin{align}
S^{(2)}_T =\int d^4x a^3 M^2 &\left[\mathcal{L}_{EH}-(3H^2+\dot{H})\frac{1}{8}\left(h^2-2h_{\mu\nu}h^{\mu\nu}\right) \right. \nonumber \\
& \left. +\frac{d\log{M^2}}{d\log{a}}\mathcal{L}_+\right]+S_{M,\delta\varphi}^{(2)}, \label{Sfinalcosmology}
\end{align}
where we have defined the lagrangian $\mathcal{L}_{EH}$ to be the quadratic Taylor expansion in the metric perturbations of the Einstein Hilbert lagrangian $\frac{1}{2}\sqrt{-g} R$, whilst $\mathcal{L}_+$ represents terms which are beyond general relativity. $S_{M,\delta\varphi}^{(2)}$ represents those terms in the quadratic matter action (given by eq.~(\ref{Smattercosmology})) that depend on $\delta\varphi$. Those terms which are quadratic in $h_{\mu\nu}$ have been incorporated into the other terms of eq.~(\ref{Sfinalcosmology}). The Lagrangians $\mathcal{L}_{EH}$ and $\mathcal{L_+}$ are given by:
\begin{align}
  \mathcal{L}_{EH}= &\frac{1}{8}\bar{\nabla}_\mu h \bar{\nabla}^\mu h - \frac{1}{4}\bar{\nabla}_\mu h^{\mu\nu}\bar{\nabla}_\nu h-\frac{1}{8}\bar{\nabla}_\mu h^{\mu\lambda}\bar{\nabla}_\nu h^{\nu}_{\,\lambda}\nonumber \\ &+\frac{1}{4}\bar{\nabla}_\mu h_{\nu\lambda}\bar{\nabla}^\nu h^{\mu\lambda}+\frac{1}{4}h^{\mu\rho}(h^{\nu\sigma} \bar{R}_{\rho\nu\mu\sigma}-h^{\nu\rho}\bar{R}_{\mu\nu})\nonumber\\
  &+\frac{1}{16}\bar{R}(h^2-2h_{\mu\nu}h^{\mu\nu})+\frac{1}{4}\bar{R}_{\mu\nu}(2h^\mu_{\,\sigma}h^{\sigma\nu}-hh^{\mu\nu}),
  \label{EHaction}
  \end{align}
  and

  \begin{widetext}
  \begin{align}
   \mathcal{L}_+=&-\frac{1}{2}H^2h_{\mu\nu}h^{\mu\nu}-\frac{1}{8}h^2-2H^2h_{\mu}^{\,\sigma}h_{\nu\sigma}u^\mu u^\nu -\frac{3}{4}H^2hh_{\mu\nu}u^\mu u^\nu-\frac{7}{4}H^2h_{\mu\nu}h_{\sigma\lambda}u^\mu u^\nu u^\sigma u^\lambda -\frac{1}{8}\bar{\nabla}_\mu h^{\sigma\lambda}\bar{\nabla}_\nu h_{\sigma\lambda}u^\mu u^\nu \nonumber\\
  & -\frac{1}{4}H h^{\nu}_{\,\mu}u^\mu\bar{\nabla}_\nu h +\frac{1}{8}\bar{\nabla}_\mu h \bar{\nabla}_\nu h u^\mu u^\nu +\frac{1}{8}\bar{\nabla}_\mu h \bar{\nabla}^\mu h -\frac{1}{2}Hh^{\nu\sigma}u^\mu\bar{\nabla}_\sigma h_{\mu\nu} -\frac{1}{4}\bar{\nabla}_\mu h \bar{\nabla}^\sigma h_{\nu\sigma}u^\mu u^\nu+\frac{1}{4}\bar{\nabla}_\mu h^{\mu\nu}\bar{\nabla}^\sigma h_{\nu\sigma} \nonumber\\
  & -\frac{1}{4}\bar{\nabla}_\mu h \bar{\nabla}_\nu h^{\mu\nu}
  -\frac{1}{2}H h_{\mu\rho}u^\mu u_\nu u^\sigma \bar{\nabla}_\sigma h^{\nu\rho} -\frac{1}{4}Hh_{\mu\nu}u^\mu u^\nu u^\sigma \bar{\nabla}_\sigma h -\frac{1}{4}u^\mu u^\nu \bar{\nabla}_\nu h_{\mu\sigma} \bar{\nabla}^\sigma h +\frac{1}{4}u^\mu u^\nu \bar{\nabla}_\sigma h \bar{\nabla}^\sigma h_{\mu\nu}\nonumber\\
  &-\frac{1}{8}\bar{\nabla}_\sigma h_{\mu\nu}\bar{\nabla}^\sigma h^{\mu\nu}
  -\frac{3}{4}Hh_{\mu\rho}u^\mu u^\nu u^\sigma \bar{\nabla}^\rho h_{\nu\sigma}+\frac{1}{4}u^\mu u^\nu \bar{\nabla}^\sigma_{\mu\sigma}\bar{\nabla}^\rho h_{\nu\rho}+\frac{1}{4}Hh_{\mu\nu}u^\mu u^\nu u^\sigma \bar{\nabla}^\rho h_{\sigma\rho}+\frac{1}{2}u^\mu u^\nu \bar{\nabla}_\nu h_{\mu\sigma}\bar{\nabla}_\rho h^{\sigma\rho} \nonumber\\
  &-\frac{1}{4}u^\mu u^\nu \bar{\nabla}_\sigma h_{\mu\nu}\bar{\nabla}_\rho h^{\sigma\rho}
  -\frac{1}{4}u_\mu u^\nu\bar{\nabla}_\rho h_{\nu\sigma}\bar{\nabla}^\rho h^{\mu\sigma}-\frac{1}{2}Hh_{\mu\nu}u^\mu u^\nu u^\sigma u^\rho u^\lambda \bar{\nabla}_\lambda h_{\sigma\rho}+\frac{1}{4}u^\mu u^\nu u^\sigma u^\rho \bar{\nabla}_\sigma h_{\mu\nu}\bar{\nabla}^\lambda h_{\rho\lambda}\nonumber\\
  &-\frac{1}{4}u^\mu u^\nu u^\sigma u^\rho\bar{\nabla}_\rho h_{\sigma\lambda}\bar{\nabla}^\lambda h_{\mu\nu},
\end{align}
\end{widetext}
 where $\bar{R}$, $\bar{R}_{\mu\nu}$ and $\bar{R}_{\rho\nu\mu\sigma}$ are the Ricci scalar, Ricci tensor and Riemann tensor for the background metric, respectively. We note that contrary to \cite{Lagos:2016wyv} these actions are explicitly written in a covariant form and, under the standard SVT decomposition, they give the evolution of scalar, vector and tensor perturbations all at once. As we will see next, this action propagates one physical scalar degree of freedom (d.o.f) , and two tensor d.o.f, and therefore we associate this model with that of a massless graviton coupled to a matter scalar field. 

Let us first consider scalar perturbations. In this case, we can choose the Newtonian gauge and hence write the perturbed metric as:
\begin{equation}
h_{\mu\nu}=-2\Phi u_\mu u_\nu -2\Psi\gamma_{\mu\nu},
\end{equation}
where $\Phi$ and $\Psi$ are the two gravitational potentials describing the scalar perturbations. After making suitable integrations by parts, (\ref{Sfinalcosmology}) can be shown to be equal to the action given by eq.~(3.33) in \cite{Lagos:2016wyv}, which indeed propagates one physical scalar d.o.f. 

Next, we write the action for vector perturbations. We calculate (\ref{Sfinalcosmology}) in a gauge such that \cite{Gleyzes:2014rba} the perturbed metric takes the following form:
\begin{align}
  h_{\mu\nu}=-u_\mu\gamma^\lambda_\nu N_\lambda-u_\nu\gamma^\lambda_\mu N_\lambda,\label{vecpert}
\end{align}
where $N_\lambda$ is a transverse vector that describes the vector perturbations of the metric and satisfies $N_\mu u^\mu=\partial_i N^i=0$. The resultant action for vector perturbations is given by:
\begin{align}
  S_v^{(2)}=\int d^4x\; \frac{1}{a}\frac{M^2}{8}\left(\partial_i N_j+\partial_j N_i\right)^2.\label{GRactionVector}
\end{align}
From here we see that $N_i$ is an auxiliary field, i.e.~does not have time derivatives and hence it does not represent a physical propagating d.o.f. As expected then, there are no physical vector perturbations propagating. 

Finally, for tensor perturbations we can write the perturbed metric as:
\begin{equation}
h_{\mu\nu}=a^2 \gamma_{\mu\alpha}\gamma_{\nu\beta}e^{\alpha\beta},\label{tensorpert}
\end{equation}
where $e^{\alpha\beta}$ describes the tensor perturbations which are traceless and transverse, that is, $e^\mu{}_\mu=0$ and $\partial^ie_{ij}=0$, respectively. In this case we find the gravitational action to be:
\begin{align}
  S_t^{(2)}=\int d^4x\,a^3\frac{M^2}{8}\left((\dot{e}_{ij})^2-\frac{1}{a^2}\left(1+\frac{d\log{M^2}}{d\log{a}}\right)(\partial_k e_{ij})^2\right).\label{GRactionTensor}
\end{align}
From here we see that $e_{ij}$ is a dynamical field that in principle has 6 d.o.f, but due to the transverse and traceless conditions, it propagates only two physical d.o.f, that we associate to the two polarisations of massless graviton. 

We note that if $M$ is the Planck mass $M_{Pl}$ we recover general relativity coupled minimally to a matter scalar field. Indeed, from the final action in eq.~(\ref{Sfinalcosmology}), we see that when the mass is constant the contribution from $\mathcal{L}_+$ vanishes and the combination of $(3H^2+\dot{H})$ contributes only as an integration constant that can be interpreted as the cosmological constant. This can be explicitly shown by integrating the Friedmann equation (\ref{fried1}) and combining the resulting solution with the matter background eq.~(\ref{phiEOM}). This interpretation is consistent with the result of using the Friedmann equations \textit{a priori} to evaluate $(3H^2+\dot{H})$ with a potential-less scalar field. Thus the correct quadratic expansion of the Einstein-Hilbert action with cosmological constant (i.e.~general relativity) is recovered in the case of a constant $M$.

\section{Scalar-Tensor theories}\label{secST}

Having studied the case of a single-tensor perturbation on a cosmological background, we now construct the most general gravitational action for cosmological perturbations of a tensor and a scalar field, that leads to second order equations of motion and is linearly diffeomorphism invariant \cite{Horndeski:1974wa,Deffayet:2009wt}. We follow the covariant procedure as in the previous section, but with the addition of a gravitational scalar field $\chi$:
\begin{align}
  \chi=\bar{\chi}(t)+\delta\chi; \quad |\delta\chi|\ll |\bar{\chi}|,
\end{align}
where $\bar{\chi}$ is the background value of the field, which we assumed to be time-dependent only to comply with the global symmetries of the background, and $\delta\chi$ is a linear perturbation 
non-minimally coupled to the metric $g_{\mu\nu}$ and its perturbation, $h_{\mu\nu}$. Since we have the same homogeneous and isotropic background as in the previous section, we also use the 1+3 split of space-time with the time-like vector $u^\mu$ and the spatial metric $\gamma_{\mu\nu}$. 

We move onto step 2 and write down the most general scalar-tensor gravitational action as:
\begin{align}
  S_G^{(2)}=\int d^4x\,a^3\; &\left[\mathcal{A}^{\mu\nu\alpha\beta} h_{\mu\nu} h_{\alpha\beta}
 + \mathcal{B}^{\mu\nu\alpha\beta\delta}\bar{\nabla}_{\delta} h_{\mu\nu} h_{\alpha\beta}
\right. \nonumber\\
&\left.+ \mathcal{C}^{\mu\nu\alpha\beta\kappa\delta}\bar{\nabla}_{\kappa} h_{\mu\nu} \bar{\nabla}_{\delta} h_{\alpha\beta} +A_{\chi}(\delta\chi)^2\right. \nonumber\\
&\left.+\mathcal{A}_\chi^{\mu\nu}\delta\chi h_{\mu\nu}+\mathcal{B}_\chi^{\mu\nu\delta}h_{\mu\nu}\bar{\nabla}_\delta\delta\chi \right. \nonumber\\
&\left.+\mathcal{C}_\chi^{\mu\nu}\bar{\nabla}_\mu\chi\bar{\nabla}_\nu\chi+\mathcal{D}_\chi^{\mu\nu\delta\kappa}\bar{\nabla}_\kappa\delta\chi\bar{\nabla}_\delta h_{\mu\nu}\right], \label{ScosmologyST}
\end{align}
where the $\mathcal{A}$, $\mathcal{B}$, and $\mathcal{C}$ are the same as those given by (\ref{Atensorcosmology})-(\ref{Ctensorcosmology}). We see that we also have two new tensors describing the self-interactions of the scalar field and three for the interactions between the scalar and tensor fields. These new tensors are arbitrary functions of the background, and hence must follow the background symmetry and can be constructed solely from the projectors $u^\mu$ and $\gamma_{\mu\nu}$. Similarly as in the previous section, we proceed to write down the most general forms these five new tensors can take: 
\begin{align}
  \mathcal{A}_\chi^{\mu\nu}=& \ A_{\chi1}u^\mu u^\nu +A_{\chi2}\gamma^{\mu\nu}\\
  \mathcal{B}_\chi^{\mu\nu\delta}=& \ B_{\chi1}u^\mu u^\nu u^\delta +B_{\chi2}u^\delta\gamma^{\mu\nu}+B_{\chi3}u^\mu\gamma^{\delta\nu}\\
  \mathcal{C}_\chi^{\mu\nu}=& \ C_{\chi1}u^\mu u^\nu +C_{\chi2}\gamma^{\mu\nu}\\
  \mathcal{D}_\chi^{\mu\nu\delta\kappa}=&\ D_{\chi1}u^\mu u^\nu u^\delta u^\kappa +D_{\chi2}u^\mu u^\nu \gamma^{\kappa\delta} + D_{\chi3}u^\kappa u^\delta \gamma^{\mu\nu} \nonumber \\ & \ + D_{\chi4} u^\mu u^\kappa \gamma^{\delta\nu} + D_{\chi5}\gamma^{\mu\nu}\gamma^{\kappa\delta} + D_{\chi6}\gamma^{\mu\kappa}\gamma^{\nu\delta},
\end{align}
while $A_{\chi}$ is a scalar and hence simply considered to be free function of time. Here, each of the coefficients $A_{\chi\,n}$, $B_{\chi\,n}$, $C_{\chi\,n}$, and $D_{\chi\,n}$ are free functions of time as well. We see that we have 14 additional free functions due to the inclusion of the scalar field $\chi$. 

We now follow step 3. Similarly as before, we add matter and consider a scalar field $\varphi$ minimally coupled to the metric $g_{\mu\nu}$. The matter quadratic action is given by eq.~(\ref{Smattercosmology}) and we must impose linear diffeomorphism invariance of the total action (gravity with matter). While the metric and matter perturbations transform as in eq.~(\ref{hgaugetransformation})-(\ref{mattertransformation}) under an infinitesimal coordinate transformation, the new scalar field transforms as:
\begin{equation}
  \delta\chi \rightarrow \delta\chi + \epsilon^\mu\bar{\nabla}_\mu\bar{\chi}. \label{chigaugetransformation}
\end{equation}
The total action given by the combination of (\ref{ScosmologyST}) and the matter action (\ref{Smattercosmology}) can now be varied under the gauge transformation. As in Sections \ref{SecFP} and \ref{secGR}, we obtain a number of Noether constraints by enforcing independent terms in the Noether identities to vanish. We find 36 Noether constraints through this process, the full list of which can be found in Appendix \ref{appendixS2}. Hence, the final action only depends on four free parameters that we name $M$, $\alpha_B$, $\alpha_K$, $\alpha_T$, following the notation of \cite{Gleyzes:2014rba}. The relation between these four final parameters and the parameters in eq.~(\ref{ScosmologyST}) are given in Appendix \ref{appendixS2}. 

Analogously to the previous section, we can write the final total gauge-invariant action as the matter action plus the quadratic expansion of the Einstein-Hilbert action given in eq.~(\ref{EHaction}) plus an additional Lagrangian $\mathcal{L}_{\chi+}$ containing terms involving the perturbed scalar $\delta\chi$ and the free functions $M$, $\alpha_B$, $\alpha_K$, $\alpha_T$: 
\begin{align}
S^{(2)}_T=\int d^4x \; a^3 M^2 & \left[\mathcal{L}_{EH}-\left(3H^2+\dot{H}\right)\frac{1}{8}\left(h^2-2h_{\mu\nu}h^{\mu\nu}\right) \right. \nonumber \\ 
& + \left. \mathcal{L}_{\chi+} \right] + S_{M,\delta\varphi}^{(2)}, \label{SfinalcosmologyST}
\end{align}
where $\mathcal{L}_{EH}$ is given by (\ref{EHaction}) whilst $\mathcal{L}_{\chi+}$ is given by:
\begin{widetext}
\begin{align}
 \mathcal{L}_{\chi+}=&\frac{1}{2}H^2\left(\alpha_M-2\alpha_T\right)h_{\mu\nu}h^{\mu\nu}-\frac{1}{8}H^2\left(5\alpha_M-4\alpha_T\right)h^2 +\frac{1}{2}H^2\left(3\alpha_M-7\alpha_T\right)h_{\mu}^{\,\sigma}h_{\nu\sigma}u^\mu u^\nu \nonumber\\
 &+\frac{1}{4}H^2\left(4\alpha_B-9\alpha_M+6\alpha_T\right)hh_{\mu\nu}u^\mu u^\nu +\frac{1}{8}H\left(\alpha_K+2H\left(10\alpha_B-\alpha_M-6\alpha_T\right)\right)h_{\mu\nu}h_{\sigma\lambda}u^\mu u^\nu u^\sigma u^\lambda
  \nonumber \\ &-\frac{1}{8}\alpha_T\bar{\nabla}_\mu h^{\sigma\lambda}\bar{\nabla}_\nu h_{\sigma\lambda}u^\mu u^\nu -\frac{1}{4}H\alpha_M h^{\nu}_{\,\mu}u^\mu\bar{\nabla}_\nu h +\frac{1}{8}\alpha_T\bar{\nabla}_\mu h \bar{\nabla}_\nu h u^\mu u^\nu +\frac{1}{8}\alpha_T\bar{\nabla}_\mu h \bar{\nabla}^\mu h \nonumber\\
  &-\frac{1}{2}H\left(2\alpha_T-\alpha_M\right)h^{\nu\sigma}u^\mu\bar{\nabla}_\sigma h_{\mu\nu}
  +\frac{1}{2}H\left(\alpha_T-\alpha_M\right)hu^\mu u^\nu u^\lambda\bar{\nabla}_\lambda h_{\mu\nu}+\frac{1}{2}H\left(\alpha_T-\alpha_M\right)hu^\mu \bar{\nabla}_\nu h_{\mu\,}^\nu\nonumber\\
  &-\frac{1}{4}\alpha_T\bar{\nabla}_\mu h \bar{\nabla}^\sigma h_{\nu\sigma}u^\mu u^\nu+\frac{1}{4}\alpha_T\bar{\nabla}_\mu h^{\mu\nu}\bar{\nabla}^\sigma h_{\nu\sigma}-\frac{1}{4}\alpha_T\bar{\nabla}_\mu h \bar{\nabla}_\nu h^{\mu\nu}-\frac{1}{2}H\left(2\alpha_T-\alpha_M\right) h_{\mu\rho}u^\mu u_\nu u^\sigma \bar{\nabla}_\sigma h^{\nu\rho} \nonumber\\
  & -\frac{1}{4}H\left(2\alpha_B+\alpha_M\right)h_{\mu\nu}u^\mu u^\nu u^\sigma \bar{\nabla}_\sigma h -\frac{1}{4}\alpha_Tu^\mu u^\nu \bar{\nabla}_\nu h_{\mu\sigma} \bar{\nabla}^\sigma h +\frac{1}{4}\alpha_Tu^\mu u^\nu \bar{\nabla}_\sigma h \bar{\nabla}^\sigma h_{\mu\nu}-\frac{1}{8}\alpha_T\bar{\nabla}_\sigma h_{\mu\nu}\bar{\nabla}^\sigma h^{\mu\nu} \nonumber\\
  &-\frac{1}{4}H\left(4\alpha_T-\alpha_M\right)h_{\mu\rho}u^\mu u^\nu u^\sigma \bar{\nabla}^\rho h_{\nu\sigma}+\frac{1}{4}\alpha_Tu^\mu u^\nu \bar{\nabla}^\sigma h_{\mu\sigma}\bar{\nabla}^\rho h_{\nu\rho}+\frac{1}{4}H\left(4\alpha_B-\alpha_M+2\alpha_T\right)h_{\mu\nu}u^\mu u^\nu u^\sigma \bar{\nabla}^\rho h_{\sigma\rho}\nonumber\\
  &+\frac{1}{2}\alpha_Tu^\mu u^\nu \bar{\nabla}_\nu h_{\mu\sigma}\bar{\nabla}_\rho h^{\sigma\rho}-\frac{1}{4}\alpha_Tu^\mu u^\nu \bar{\nabla}_\sigma h_{\mu\nu}\bar{\nabla}_\rho h^{\sigma\rho}-\frac{1}{4}\alpha_Tu_\mu u^\nu\bar{\nabla}_\rho h_{\nu\sigma}\bar{\nabla}^\rho h^{\mu\sigma} \nonumber\\
  &-\frac{1}{2}H\left(\alpha_T-\alpha_B\right)h_{\mu\nu}u^\mu u^\nu u^\sigma u^\rho u^\lambda \bar{\nabla}_\lambda h_{\sigma\rho} +\frac{1}{4}\alpha_Tu^\mu u^\nu u^\sigma u^\rho \bar{\nabla}_\sigma h_{\mu\nu}\bar{\nabla}^\lambda h_{\rho\lambda} -\frac{1}{4}\alpha_Tu^\mu u^\nu u^\sigma u^\rho\bar{\nabla}_\rho h_{\sigma\lambda}\bar{\nabla}^\lambda h_{\mu\nu}\nonumber\\
& -\frac{1}{2\dot{\bar{\chi}}^4}\left(6\alpha_B\dot{H}^2\dot{\bar{\chi}}^2+\dot{H}\dot{\bar{\chi}}\left(6H\left(H\alpha_B\left(3+\alpha_M\right)+\dot{\alpha}_B\right)\dot{\bar{\chi}}-\alpha_K\ddot{\bar{\chi}}\right)\right.\nonumber\\
  &\left.-H\left(-6\alpha_B\dot{\bar{\chi}}^2\ddot{H}-2\alpha_K\ddot{\bar{\chi}}^2+\dot{\bar{\chi}}\left(\left(H\alpha_K\left(3+\alpha_M\right)+\dot{\alpha}_K\right)\ddot{\bar{\chi}}+\alpha_K\dddot{\bar{\chi}}\right)\right)\right)\left(\delta\chi\right)^2\nonumber\\
  &+\frac{1}{2\dot{\bar{\chi}}^2}\left(\alpha_B\dot{H}+H\left(H\left(\alpha_B-\alpha_M+\alpha_B\alpha_M+\alpha_T\right)+\dot{\alpha}_B\right)\right)\bar{\nabla}_\mu\delta\chi\bar{\nabla}^\mu\delta\chi\nonumber\\
  &+\frac{1}{2\dot{\bar{\chi}}^2}\left(2\alpha_B\dot{H}+H\left(\alpha_K+2H\left(\alpha_B-\alpha_M+\alpha_B\alpha_M+\alpha_T\right)+2\dot{\alpha}_B\right)\right)u^\mu u^\nu \bar{\nabla}_\mu\delta\chi \bar{\nabla}_\nu\delta\chi\nonumber\\
  &+\frac{1}{2  \dot{\bar{\chi}}^2} u^\mu   u^\nu   u^\rho   h_{\nu\rho}   \bar{\nabla}_\mu \delta\chi  \left( \dot{\bar{\chi}}  \left(H \left(4 \dot{ \alpha_B }+ \alpha_K +2 H (2  \alpha_B   \alpha_M - \alpha_B - \alpha_M + \alpha_T )\right)+4  \alpha_B  \dot{H}\right)-2 \alpha_B  H \ddot{\bar{\chi}}\right)\nonumber\\
    &-\frac{1}{2  \dot{\bar{\chi}}}H  u^\mu   u^\nu  (2\alpha_B - \alpha_M + \alpha_T )  \bar{\nabla}_\mu \delta\chi \bar{\nabla}_\nu  h +\frac{1}{2  \dot{\bar{\chi}}}H  u^\mu   u^\nu  (4\alpha_B - \alpha_M + \alpha_T )  \bar{\nabla}^\rho  \delta\chi  \bar{\nabla}_\nu  h_{\mu\rho}\nonumber\\
    & -\frac{1}{2  \dot{\bar{\chi}}}H  u^\mu   u^\nu  (2  \alpha_B - \alpha_M + \alpha_T )  \bar{\nabla}^\rho  \delta\chi  \bar{\nabla}_\rho h_{\mu\nu} +\frac{1}{2\dot{\bar{\chi}}}H  u^\mu   u^\nu  ( \alpha_T - \alpha_M )  \bar{\nabla}_\mu \delta\chi   \bar{\nabla}_\rho h_\nu^\rho+\frac{1}{2  \dot{\bar{\chi}}}H ( \alpha_M - \alpha_T )  \bar{\nabla}^\mu  \delta\chi  \bar{\nabla}_\mu  h\nonumber\\
    & +\frac{1}{2  \dot{\bar{\chi}}}H ( \alpha_T - \alpha_M )\bar{\nabla}^\mu  \delta\chi   \bar{\nabla}_\nu  h_\mu^\nu +\frac{1}{ \dot{\bar{\chi}}} u^\mu   h_{\mu\nu}  \bar{\nabla}^\nu \delta\chi  \left(H \left(2 \dot{ \alpha_B }+H (2\alpha_B  ( \alpha_M +1)- \alpha_M + \alpha_T )\right)+2  \alpha_B  \dot{H}\right)-\frac{ \alpha_B  H \ddot{\bar{\chi}}  u^\mu   h   \bar{\nabla}_\mu \delta\chi }{ \dot{\bar{\chi}}^2}\nonumber\\
    &-\frac{1}{2 \dot{\bar{\chi}}^2} \delta\chi   u^\mu   u^\nu   h_{\mu\nu}  \left(2  \dot{C}_{\chi2}   \dot{\bar{\chi}}^3+2  \dot{\bar{\chi}}^2 ( \ddot{D}_{\chi2} +2  \dot{D}_{\chi5}   H)+\ddot{\bar{\chi}} \left(H \left(2 \dot{ \alpha_B }+ \alpha_K +2 H ( \alpha_B   ( \alpha_M +2)- \alpha_M + \alpha_T )\right)+2  \alpha_B  \dot{H}\right)\right.\nonumber\\
    &\left.-2 H \dot{\bar{\chi}} \left(\dot{H} (5  \alpha_B - \alpha_M + \alpha_T )+H\left(\dot{ \alpha_B }+ \alpha_B   \alpha_M  H\right)\right)\right)-\frac{1}{ \dot{\bar{\chi}}^2} \delta\chi   h  \left( \dot{C}_{\chi2}   \dot{\bar{\chi}}^3+ \dot{\bar{\chi}}^2 ( \ddot{D}_{\chi2}+2  \dot{D}_{\chi5}  H)-H  \dot{\bar{\chi}} \left(\dot{H} (2  \alpha_B - \alpha_M + \alpha_T )\right.\right.\nonumber\\
    &\left.\left.+H \left(\dot{ \alpha_B }+ \alpha_B   \alpha_M 
   H\right)\right)+\ddot{\bar{\chi}} \left(H \left(\dot{ \alpha_B }+H ( \alpha_B 
   ( \alpha_M +5)- \alpha_M + \alpha_T )\right)+ \alpha_B 
   \dot{H}\right)\right)\label{Lchiplus}
\end{align}
\end{widetext}
Once again, having obtained a form for the fully covariant diffeomorphism invariant action, we can study the scalar, vector, and tensor actions separately. As expected, we will see that this action propagates two physical scalar d.o.f (one from matter and one from the gravitational scalar $\chi$), in addition to two tensor d.o.f.

The scalar action can be calculated by expressing the metric perturbation $h_{\mu\nu}$ in terms of the standard cosmological perturbation variables $\Phi$, $\Psi$, $E$, and $B$. After making suitable integrations by parts and the field redefinition of the perturbed gravitational scalar $ \delta\chi\rightarrow\bar{\chi} \delta\chi$, the action in eq.~(\ref{SfinalcosmologyST}) can be shown to be equal to the action given by eq.~(4.15) in \cite{Lagos:2016wyv}, which indeed propagates two physical scalar d.o.f. 

For vector perturbations, we again use the decomposition given by (\ref{vecpert}) in the unitary gauge where $\delta\chi=0$. The resultant action for vector perturbations is then given by:
\begin{align}
  S_v^{(2)}=\int d^4x\frac{1}{a}\frac{M^2}{8}\left(\partial_i N_j+\partial_j N_i\right)^2.\label{STactionVector}
\end{align}
As in the previous section, this action does not propagate any physical vector perturbations. 

Finally, for tensor perturbations, using the decomposition given by (\ref{tensorpert}), we find that the resultant action is given by:
\begin{align}
  S_t^{(2)}=\int d^4x\,a^3\frac{M^2}{8}\left((\dot{e}_{ij})^2-\frac{(1+\alpha_T)}{a^2}(\partial_k e_{ij})^2\right).\label{STactionTensor}
\end{align}
which propagates two physical d.o.f. We emphasise that both of the results (\ref{STactionVector}) and (\ref{STactionTensor}) are in agreement with those of \cite{Gleyzes:2014rba,Bellini:2014fua}.

The results of Section \ref{secGR} can be recovered by setting the three $\alpha$ parameters to the following values:
\begin{equation}
  \alpha_B=0, \; \alpha_K=0, \; \alpha_T=\frac{d\log{M^2}}{d\log{a}}.
\end{equation} 
In this case we get that 
\begin{equation}
 \mathcal{L}_{\chi+}=\frac{d\log{M^2}}{d\log{a}}\mathcal{L}_+,
\end{equation}
and we recover the Friedmann equation (\ref{fried1}) from the Noether constraint given in eq.~(\ref{friedST}), and as result all the interactions between the metric and $\delta\chi$ and the self-interactions of $\delta\chi$ vanish.

Now that we have identified the relevant free parameters characterising scalar-tensor gravity theories, it is possible to identify the physical effects of each one of them, and constrain them with observations. Indeed, as explained in \cite{Bellini:2014fua} $M$ is a generalised Planck mass, $\alpha_T$ induces a tensor speed excess, $\alpha_K$ is a kineticity term determining the kinetic energy of the gravitational scalar $\delta \chi$, and $\alpha_B$ is a braiding term that induces kinetic mix between the perturbed metric and gravitational scalar, and thus contributes to the kinetic energy of $\delta\chi$ indirectly, through backreaction with gravity. All these parameters can be constrained with current cosmological data with numerical codes such as the ones given in \cite{Zumalacarregui:2016pph,EFTCAMB1}, and hence we can find the family of such modified gravity theories that are compatible with cosmological data. As expected, observational constraints are compatible with the GR values of $\alpha_T=\alpha_K=\alpha_B=0$ and constant $M$, and strongly disfavour large deviations from these values \cite{HiCLASS}.


\section{Vector-Tensor theories}\label{secVT}

In this section we construct the most general diffeomorphism invariant quadratic action for linear cosmological perturbations when the gravitational fields are given by a metric $g_{\mu\nu}$ and a vector field $\zeta^\mu$ (see \cite{Jacobson:2000xp,Zlosnik:2006zu} for Einstein-Aether and generalized Einstein-Aether and \cite{Heisenberg:2014rta} for generalized Proca theories). Similarly as the previous sections, we add a matter scalar field $\varphi$ minimally coupled to the metric only. 

We again follow the covariant procedure for a homogeneous and isotropic background. The perturbed metric and matter field are given by eq.~(\ref{TensorMatterLinearPerts}), and now we add a gravitational vector field:
\begin{align}
  \zeta^\mu=\bar{\zeta}(t)u^\mu+\delta\zeta^\mu; \quad |\delta\zeta^\mu|\ll|\bar{\zeta} u^\mu|
\end{align}
where $\bar{\zeta}$ is the background value of the field, which must be a function of time only and proportional to $u^\mu$ due to the symmetries of the background. Here, $\delta\zeta^\mu$ is a linear perturbation and is a function of space and time. 

We now follow step 2 and, due to the global background symmetries, we use the 3+1 split of space-time as in the previous sections, giving us the time-like vector $u^\mu$ and the spatial metric $\gamma_{\mu\nu}$ as the projectors to describe our background metric. We write the most general gravitational action which is quadratic in the perturbation fields $h_{\mu\nu}$ and $\delta\zeta^\mu$ and lead to a maximum of second-order derivatives in the equations of motion:
\begin{align}
    S_G^{(2)}&=\int d^4x\,a^3\;\left[\mathcal{A}^{\mu\nu\alpha\beta} h_{\mu\nu} h_{\alpha\beta}
 + \mathcal{B}^{\mu\nu\alpha\beta\delta}\bar{\nabla}_{\delta} h_{\mu\nu} h_{\alpha\beta}
\right.\nonumber\\
&+ \mathcal{C}^{\mu\nu\alpha\beta\kappa\delta}\bar{\nabla}_{\kappa} h_{\mu\nu} \bar{\nabla}_{\delta} h_{\alpha\beta} +\mathcal{A}_{\zeta^2}^{\mu\nu}\delta\zeta_\mu\delta\zeta_\nu\nonumber\\
&+\mathcal{A}_{h\zeta}^{\mu\nu\lambda}\delta\zeta_\lambda h_{\mu\nu}
+B_{\zeta}u^\mu\gamma^{\nu\lambda}\delta\zeta_\mu\bar{\nabla}_\nu\delta\zeta_\lambda\nonumber\\
& +\mathcal{B}_{\zeta}^{\mu\nu\lambda\kappa}h_{\mu\nu}\bar{\nabla}_\kappa\delta\zeta_\lambda +\mathcal{C}_{\zeta}^{\mu\nu\kappa\delta}\bar{\nabla}_\kappa\delta\zeta_\mu\bar{\nabla}_\delta\delta\zeta_\nu\nonumber\\
&\left.+\mathcal{D}_{\zeta}^{\mu\nu\lambda\kappa\delta}\bar{\nabla}_\kappa\delta\zeta_\lambda\bar{\nabla}_\delta h_{\mu\nu}\right]. \label{ScosmologyVT}
\end{align}
The tensors $\mathcal{A}$, $\mathcal{B}$, and $\mathcal{C}$ are the same as those given by (\ref{Atensorcosmology})-(\ref{Ctensorcosmology}). The new tensors describing self interactions of the vector field and interactions between the vector and tensor fields are given by:
\begin{widetext}
\begin{align}
  \mathcal{A}_{\zeta^2}^{\mu\nu}=&A_{\zeta1}u^\mu u^\nu + A_{\zeta2}\gamma^{\mu\nu}\\
  \mathcal{A}_{h\zeta}^{\mu\nu\lambda}=&A_{\zeta3}u^\mu u^\nu u^\lambda +A_{\zeta4}\gamma^{\mu\nu}u^\lambda+A_{\zeta5}u^\mu\gamma^{\nu\lambda}\\
  \mathcal{B}_{\zeta}^{\mu\nu\lambda\kappa}=&B_{\zeta1}u^\mu u^\nu u^\lambda u^\kappa +B_{\zeta2}u^\mu u^\nu\gamma^{\lambda\kappa}+B_{\zeta3}u^\kappa u^\lambda \gamma^{\mu\nu}+B_{\zeta4}u^\mu u^\lambda\gamma^{\kappa\nu}+B_{\zeta5}u^\mu u^\kappa\gamma^{\nu\lambda}+B_{\zeta6}\gamma^{\mu\nu}\gamma^{\lambda\kappa} +B_{\zeta7}\gamma^{\mu\kappa}\gamma^{\nu\lambda}\\
  \mathcal{C}_\zeta^{\mu\nu\kappa\delta}=&C_{\zeta1}u^\mu u^\nu u^\kappa u^\delta +C_{\zeta2}u^\mu u^\nu \gamma^{\kappa\delta} +C_{\zeta3}u^\kappa u^\delta\gamma^{\mu\nu} +C_{\zeta4}u^\mu u^\kappa\gamma^{\nu\delta}+C_{\zeta5}\gamma^{\mu\nu}\gamma^{\kappa\delta}+C_{\zeta6}\gamma^{\mu\kappa}\gamma^{\nu\delta}\\
  \mathcal{D}_{\zeta}^{\mu\nu\lambda\kappa\delta}=&D_{\zeta1}u^\mu u^\nu u^\lambda u^\kappa u^\delta+D_{\zeta2}u^\lambda u^\kappa u^\delta \gamma^{\mu\nu} + D_{\zeta3}u^\lambda u^\mu u^\nu \gamma^{\kappa\delta} +D_{\zeta4}u^\lambda u^\mu u^\kappa \gamma^{\delta\nu} +D_{\zeta5}u^\mu u^\nu u^\delta \gamma^{\kappa\lambda} +D_{\zeta6}u^\mu u^\kappa u^\delta \gamma^{\nu\lambda} \nonumber\\
  & + D_{\zeta7}u^\lambda\gamma^{\mu\nu}\gamma^{\kappa\delta} + D_{\zeta8}u^\lambda\gamma^{\mu\kappa}\gamma^{\delta\nu} +D_{\zeta9}u^\mu\gamma^{\nu\kappa}\gamma^{\delta\lambda}+D_{\zeta10}u^\kappa\gamma^{\mu\nu}\gamma^{\delta\lambda} +D_{\zeta11}u^\kappa\gamma^{\mu\delta}\gamma^{\nu\lambda}+D_{\zeta12}u^\mu\gamma^{\nu\lambda}\gamma^{\kappa\delta},
\end{align}
\end{widetext}
with each of the coefficients $A_{\zeta\,n}$, $B_{\zeta\,n}$, $C_{\zeta\,n}$, and $D_{\zeta\,n}$ being free functions of time, in addition to the scalar $B_{\zeta}$. We note that we have gained 31 additional free functions due to the inclusion of the vector field $\zeta^\mu$.

We now follow step 3. As before, the total action is given by the combination of the quadratic gravitational action in eq.~(\ref{ScosmologyVT}) and the quadratic matter action in eq.~(\ref{Smattercosmology}). We must now impose linear diffeomorphism invariance of the total action under a coordinate transformation given by (\ref{coordtransformation}). The metric and matter perturbations transforming as in eq.~(\ref{hgaugetransformation})-(\ref{mattertransformation}), whilst the new vector field transforms as:
\begin{align}
  \delta\zeta^\mu\rightarrow& \delta\zeta^\mu+\bar{\zeta} u^\nu\bar{\nabla}_\nu\epsilon^\mu-\epsilon^\nu\bar{\nabla}_\nu(\bar{\zeta} u^\mu).\label{vectortransformation}
\end{align}
As in the previous sections, by varying the total action under infinitesimal gauge transformations, we obtain four Noether identities which in turn lead to a number of Noether constraints. For vector-tensor theories we obtain 47 Noether constraints, the full list of which can be found in Appendix \ref{appendixVT}. Therefore, the final action has 10 free parameters (arbitrary functions of time): $C_{\zeta2}$-$C_{\zeta6}$, $D_{\zeta4}$, $D_{\zeta7}$, $D_{\zeta9}$, and $C_5$. Similarly as before, we define an effective mass of the vector-tensor action and its running as:
\begin{equation}
  M^2=-8C_5+2(C_{\zeta5}+C_{\zeta6})\bar{\zeta}^2, \quad \alpha_M=\frac{d\log{M^2}}{d\log{a}}.
  \label{VTmass}
\end{equation}
Analogously to (\ref{Sfinalcosmology}) and (\ref{SfinalcosmologyST}), the total final gauge-invariant action can be written as the matter action plus the quadratic expansion of the Einstein-Hilbert action (\ref{EHaction}) plus an additional Lagrangian $\mathcal{L}_{\zeta+}$ containing terms involving the perturbed vector $\delta\zeta^\mu$ and involving the 10 free parameters: 
\begin{align}
S^{(2)}_T=\int d^4x a^3 M^2 &\left[\mathcal{L}_{EH}-(3H^2+\dot{H})\frac{1}{8}\left(h^2-2h_{\mu\nu}h^{\mu\nu}\right) \right. \nonumber \\ &\left.+\mathcal{L}_{\zeta+}\right]+S^{(2)}_{M,\delta\varphi}.\label{SfinalcosmologyVT}
\end{align}
We do not present $\mathcal{L}_{\zeta+}$ in its entirety here due to the excessive length of the expression. However, having obtained a form for the fully covariant diffeomorphism invariant action, we can study the actions for scalar, vector, and tensor type perturbations separately. For scalar perturbations, we proceed as in the previous sections and, in addition, decompose the vector perturbation $\delta\zeta^\mu$:
\begin{align}
  \delta\zeta^\mu=\left(Z_0-\bar{\zeta}\Phi\right)u^\mu+\gamma^{\mu\nu}\partial_\nu Z_1,\label{ScalarVTdecomp}
\end{align}
where $Z_0$ and $Z_1$ are the two scalar perturbations. We decompose the time perturbation of the vector field as in eq.~(\ref{ScalarVTdecomp}) so that the field $\Phi$ appears explicitly as an auxiliary field (i.e.~without that derivatives) in the final action for scalar perturbations. 
After making suitable integrations by parts, (\ref{SfinalcosmologyVT}) can be shown to be equal to the following action:
\begin{widetext}
\begin{align}
S^{(2)}_{s}=\int d^4x\,a^3&\left(T_{\Phi^2}\Phi^2+T_{\partial^2\Phi^2}\partial_i\Phi\partial^i\Phi+T_{\dot{\Psi}^2}\dot{\Psi}^2+T_{\partial^2\Psi^2}\partial_i\Psi\partial^i\Psi+T_{\Phi\Psi}\Phi\Psi+T_{\Phi\dot{\Psi}}\Phi\dot{\Psi} 
 +T_{\partial\Phi\partial\Psi}\partial_i\Phi\partial^i\Psi+T_{Z_0\Phi}Z_0\Phi\right.\nonumber\\* &+T_{\dot{Z}_0\Phi}\dot{Z}_0\Phi 
 +T_{\partial Z_0\partial\Phi}\partial_i Z_0\partial^i\Phi+T_{Z_0\Psi}Z_0\Psi+T_{\dot{Z}_0\Psi}\dot{Z}_0\Psi+T_{\dot{Z}_0\dot{\Psi}}\dot{Z}_0\dot{\Psi}
 +T_{\partial Z_0\partial\Psi}\partial_i Z_0\partial^i\Psi+T_{Z_0^2}Z_0^2+T_{\dot{Z}_0^2}\dot{Z}_0^2 \nonumber\\*
&+T_{\partial^2Z_0^2}\partial_i Z_0\partial^i Z_0 +T_{\partial\dot{Z}_0\partial Z_1}\partial_i\dot{Z}_0\partial^i Z_1 +T_{\partial Z_0 \partial Z_1}\partial_i Z_0 \partial^i Z_1 
+T_{\partial\Phi\partial Z_1}\partial_i\Phi\partial^i Z_1+T_{\partial\Phi\partial\dot{Z}_1}\partial_i\Phi\partial^i\dot{Z}_1+T_{\dot{\Psi}\dot{Z}_1}\dot{\Psi}\dot{Z}_1 \nonumber\\*
&\left.+T_{\partial\Psi\partial Z_1}\partial_i\Psi\partial^i Z_1+T_{\partial\dot{\Psi}\partial Z_1}\partial_i\dot{\Psi}\partial^i Z_1+T_{\partial^2 Z_1^2}\partial_i Z_1 \partial^i Z_1 + T_{\partial^2 \dot{Z}_1^2}\partial_i \dot{Z}_1 \partial^i \dot{Z}_1 +T_{\partial^4 Z_1^2}\partial_i\partial_j Z_1 \partial^i\partial^j Z_1\right)\label{VTscalar}
\end{align}
\end{widetext}
where we have defined 27 auxiliary coefficients $T$, one for each interaction term, which are not independent and instead can all be expressed in terms of the 10 free parameters. We do not give the explicit form of the $T$ coefficients here for brevity's sake, but a dictionary relating the $T$ parameters to the 10 remaining free functions is given in Appendix \ref{appendixVT2}. As shown in Appendix \ref{appendixVT2}, this action for scalar perturbations in fact depends only on 9 different combinations of the 10 free parameters, and hence by observing the cosmological effect of these perturbations we can only constrain these 9 combinations. We also note that the result found here is a subclass of that one in \cite{Lagos:2016wyv}, where it was found that the action for scalar perturbations depended indeed on 10 combinations of free parameters instead of 9. The difference lies in the fact that in this paper we have imposed gauge invariance of the full covariant fields, and hence for scalar {\it and} vector perturbations (tensor perturbations are gauge independent in this case), whereas in \cite{Lagos:2016wyv} the gauge invariance was imposed only on scalar perturbations. Vector perturbations lead to additional Noether constraints, one of which relates the 10 free parameters found in \cite{Lagos:2016wyv} and thus reduces the action for scalar perturbations to the one presented in this paper.

As explained in \cite{Lagos:2016wyv}, one might naively expect this action to propagate at most two physical scalar d.o.f, namely, a potential helicity-0 mode of a massive vector and the matter field. However, we find that this action propagates three physical scalar d.o.f described by $\Psi$, $Z_0$ and $Z_1$, while $\Phi$ is an auxiliary field. Since we do not have the full non-linear completion of these models, at this level it is not possible to identify with certainty where the third scalar comes from, but it is likely to represent an unstable mode called Boulware-Deser ghost \cite{Boulware:1973my}, which can typically appear in modified gravity theories unless the field interactions are restricted to particular forms (see for instance \cite{deRham:2010ik,Heisenberg:2014rta,Horndeski:1974wa,Crisostomi:2016czh}). In our case, it is possible to avoid such mode, by an appropriate choice of the free parameters. Indeed, inspired by the Generalised Proca theory \cite{Heisenberg:2014rta}, which describes the ghost-free action of a massive vector field non-minimally coupled to a metric, we can fix two free parameters, namely $C_{\zeta2}=0$ and $D_{\zeta4}=\bar{\zeta}C_{\zeta4}$, and make the kinetic terms of $Z_0$ vanish, in which case $Z_0$ becomes an auxiliary field and action (\ref{VTscalar}) propagates only two physical scalar d.o.f described by $\Psi$ and $Z_1$.

For vector perturbations, we use the decomposition given by (\ref{vecpert}) for $h_{\mu\nu}$. The following decomposition of $\delta\zeta^\mu$ into a divergence-less spatial vector $Z^\mu$ is used:
\begin{align}
\delta\zeta^\mu=\left(\delta Z_\nu-\bar{\zeta}N_\nu\right)\gamma^{\mu\nu},
\end{align}
where we have decomposed the vector perturbation in this way so that the field $N_\nu$ appears explicitly as an auxiliary field in the final action.
The resultant action for vector perturbations is given by:
\begin{widetext}
\begin{align}
  S_v^{(2)}=\int d^4x\,a^3&\left[ T_{\partial^2 N^2}\frac{1}{a^4}\partial_i N_j \partial_i N_j +T_{\partial N \partial \delta Z}\frac{1}{a^4}\partial_i N_j \partial_i \delta Z_j+T_{\delta Z^2}\delta Z_i \delta Z^i + T_{\delta\dot{Z}^2}\delta \dot{Z}_i\delta\dot{Z}^i+T_{\partial^2\delta Z^2}\frac{1}{a^4}\partial_i \delta Z_j \partial_i \delta Z_j\right].\label{VTactionVector}
\end{align}
\end{widetext}
Again, we have defined intermediate coefficients $T$, one for each interaction term, and they are related to the 10 free parameters ad shown in Appendix \ref{appendixVT3}. In the appendix we also show that there are only 5 independent combinations of the free parameters this action depends on, and hence, if vector perturbations leave any signature, we can at best constrain these 5 combinations. We note that this action has two vector fields $N_i$ and $\delta Z_i$, though $N_i$ appears as an auxiliary field (i.e. without time derivatives). Thus there is only one physical vector field propagating, which has two d.o.f that we associate to two helicities $\pm 1$ of a vector field. 

For tensor perturbations, again using the decomposition given by (\ref{tensorpert}), we find the following action:
\begin{align}
  S_t^{(2)}=\int d^4x\,a^3\frac{M^2}{8}\left((\dot{e}_{ij})^2-\frac{(1+\alpha_T)}{a^2}(\partial_k e_{ij})^2\right),\label{VTactionTensor}
\end{align}
which propagates one physical tensor field, and hence two d.o.f. Here we have defined the parameter $\alpha_T$ to represent the speed excess of gravitational waves in vector-tensor gravity, analogously to eq.~(\ref{STactionTensor}):
\begin{align}
 \alpha_T&=\alpha_M-\frac{1}{HM^2}\left(4\dot{C}_{\zeta5}\bar{\zeta}^2 + 4C_{\zeta5}\bar{\zeta}\left(H\bar{\zeta}+2\dot{\bar{\zeta}}\right)+2\dot{D}_{\zeta9}\bar{\zeta} \right.\nonumber\\
& \left. - 2D_{\zeta7}\dot{\bar{\zeta}}+2D_{\zeta9}\left(H\bar{\zeta}+\dot{\bar{\zeta}}\right)\right).\label{VTalphaT}
\end{align}
We emphasise that the results shown in this section for vector and tensor perturbations have never been shown before, as the work on \cite{Lagos:2016wyv} focused only on scalar perturbations. Here we have found that the structure of the action for tensor perturbations is the same as that one for scalar-tensor theories as there are no extra tensor fields involved in the model, and can help constrain only one of the 10 free parameters of the model. However, the action for vector perturbations is quite different to that one for scalar-tensor theories, as vector-tensor theories propagate one physical vector field, whose potential signatures could help constrain at most 5 of the 10 free parameters of the model. 

Finally, we mention that the results of Section \ref{secGR} can be recovered by setting the free parameters to the following values:
\begin{align}
C_{\zeta2}=&C_{\zeta2}=C_{\zeta3}=C_{\zeta4}=C_{\zeta5}=C_{\zeta6}=0\nonumber\\
D_{\zeta4}=&D_{\zeta7}=D_{\zeta9}=0.
\end{align}
In which case, we find that:
\begin{align}
  M^2=&-8C_5, \; \alpha_T=\alpha_M, \; \mathcal{L}_{\zeta+}=\alpha_M\mathcal{L}_+.
\end{align} 

Similarly to the case of scalar-tensor theories, it should be possible to identify the physical effects associated to each one of the 10 free parameters found for vector-tensor theories, in addition to theoretical and numerical constraints on them. Up to date, such analysis has been done for the special case of Generalised Proca theories \cite{DeFelice:2016yws,DeFelice:2016uil,Nakamura:2017dnf,deFelice:2017paw}, but the parametrised action presented in this section is more general and its further analysis will be left for future work.


\section{The role of global symmetries and the number of free parameters}\label{Symmetries}

For each case we have considered, we have found that a finite number of free constants or functions can parametrise a general class of linearized theories. The number of free functions, or parameters, that come out of the Noether constraints will depend on the field content but also, crucially, on what we are assuming about the background space-time. So, for example, we found that on Minkowski space, the result was one free constant which lead us to the massless Fierz-Pauli action -- linearized general relativity. But when we repeated the calculation on an expanding background and assumed a $3+1$ split, we found one free {\it function} of time, $M^2(t)$ which is more general than the linearized general relativity on an expanding background. To understand why this is so, we need to look at the role that global symmetries play in the procedure.
 
Global symmetries are symmetries of the background, which are independent of the local gauge symmetries of the perturbations; we can have gauge invariance around a background with any symmetry. For instance, as shown in \cite{Lagos:2016wyv}, we can have linearly diffeomorphism invariant actions in Lorentz-breaking theories such as Einstein-Aether. For this reason, we must enforce both types of symmetries -- gauge and global -- {\it independently}. While from the covariant action approach it is clear that Noether constraints enforce gauge symmetries, we clarify that making the appropriate choice of background projectors to construct the most general action in step 2 will enforce global symmetries of the background. 

\subsection{Considering Minkowski again}
In order to illustrate this key point, let us go back to the quadratic action for a single tensor (or metric) in Minkowski space-time, shown in Section \ref{SecFP}. We found that if we construct the most general action in step 2 using the a single projector $\eta_{\mu\nu}$ the final diffeomorphism invariant action does not have any free parameters. Let us now repeat the calculation but now assuming a $3+1$ split of the space-time, and hence using the projectors $u^{\mu}$ and $\gamma_{\mu\nu}$ to construct the most general action in step 2. In such a case the coefficients $\mathcal{A}$ and $\mathcal{B}$ in eq.~(\ref{GRCoeffs}) would be replaced by the expressions given in eq.~(\ref{Atensorcosmology}) and (\ref{Ctensorcosmology}). At this point we can already see that there are many more free parameters in this action, compared to the six parameters in eq.~(\ref{GRCoeffs}). After following step 3, and imposing linear diffeomorphism invariance, the resulting action is the following:
\begin{widetext}
\begin{align}
S^{(2)}=&\int d^4x\,\left[\left(\frac{1}{2}\partial_\mu h \partial^\mu h - \partial_\mu h^{\mu\nu}\partial_\nu h-\frac{1}{2}\partial_\mu h_{\nu\lambda}\partial^\mu h^{\nu\lambda}+\partial_\mu h_{\nu\lambda}\partial^\nu h^{\mu\lambda}\right)\right.\nonumber\\
&\left.+\alpha \left(\frac{1}{2}u^\mu u^\nu \partial_\mu h^{\alpha\beta}\partial_\nu h_{\alpha\beta} -\frac{1}{2}u^\mu u^\nu \partial_\mu h \partial_\nu h +2u^\mu u^\nu \partial_\nu h_\mu^\lambda \partial_\lambda h - u^\mu u^\nu \partial_\lambda h_{\mu\nu} \partial^\lambda h\right.\right.\nonumber\\
&\left.\left. -u^\mu u^\nu \partial_\lambda h_\mu^\lambda \partial_\rho h_\nu^\rho -2 u^\mu u^\nu\partial_\nu h_{\mu\lambda}\partial_\rho h^{\lambda\rho}+u^\mu u^\nu\partial_\lambda h_{\mu\nu}\partial_\rho h^{\lambda\rho}+u^\mu u^\nu\partial_\rho h_{\nu\lambda}\partial^\rho h_\mu^\lambda \right)\right], \label{SrecoverFP}
\end{align}
\end{widetext}
where we have again redefined the perturbation $h_{\mu\nu}$ to eliminate an overall free factor in the action, and $\alpha$ is an arbitrary constant. Unlike the action in eq.~(\ref{SFPfinal}), here we find one free parameter $\alpha$ that multiplies terms involving the time-like vector $u^\mu$. We emphasise that to obtain this action we have indeed used that $u^\mu$ and $\gamma_{\mu\nu}$ are those of a Minkowski metric, and hence the difference in the two actions does not come from the possibility that the background metric of eq.~(\ref{SrecoverFP}) is more general than the one previously used. The difference is due to the fact that in arriving at (\ref{SrecoverFP}) we have not respected the global symmetries of the background space-time. As our background metric is Minkowski, our action cannot contain any Poincare symmetry breaking terms such as $u^{\mu}$. Therefore, to obtain the correct action for linear perturbations around Minkowski we must impose the background symmetry and enforce $\alpha$ to vanish. In this way we recover our previous result. 



We can explore this aspect even further by constructing the most general linearly diffeomorphism invariant action for a single metric around Minkiwski, but now assuming a $1+1+2$ split of the space-time. 
Thus we now define three projectors: a time-like vector $u^\mu$, a space-like vector $x^\mu$, and a space-like tensor $\gamma_{\mu\nu}$ such that the background metric is expressed as:
\begin{align}
\bar{g}_{\mu\nu}=-u_\mu u_\nu +x_\mu x_\nu +\gamma_{\mu\nu}.\label{bianchisplit}
\end{align}
Repeating the analysis of Section \ref{SecFP}, we can now construct the most general action quadratic in $h_{\mu\nu}$ with at most second order equations of motion in terms of $u^\mu$, $x^\mu$, and $\gamma_{\mu\nu}$. The resulting action is now given by:
\begin{widetext}
\begin{align}
S^{(2)}=&\int d^4x\,\left[\left(\frac{1}{2}\partial_\mu h \partial^\mu h - \partial_\mu h^{\mu\nu}\partial_\nu h-\frac{1}{2}\partial_\mu h_{\nu\lambda}\partial^\mu h^{\nu\lambda}+\partial_\mu h_{\nu\lambda}\partial^\nu h^{\mu\lambda}\right)\right.\nonumber\\
&\left.+\alpha_1 \left(\frac{1}{2}u^\mu u^\nu \partial_\mu h^{\lambda\rho}\partial_\nu h_{\lambda\rho} -\frac{1}{2}u^\mu u^\nu \partial_\mu h \partial_\nu h +2u^\mu u^\nu \partial_\nu h_\mu^\lambda \partial_\lambda h - u^\mu u^\nu \partial_\lambda h_{\mu\nu} \partial^\lambda h-u^\mu u^\nu \partial_\lambda h_\mu^\lambda \partial_\rho h_\nu^\rho -2 u^\mu u^\nu\partial_\nu h_{\mu\lambda}\partial_\rho h^{\lambda\rho} \right.\right.\nonumber\\
&\left.\left. +u^\mu u^\nu\partial_\lambda h_{\mu\nu}\partial_\rho h^{\lambda\rho}+u^\mu u^\nu\partial_\rho h_{\nu\lambda}\partial^\rho h_\mu^\lambda \right)\right. \left.+\alpha_2\left(\frac{1}{2}u^\mu x^\nu \partial_\mu h^{\lambda\rho}\partial_\nu h_{\lambda\rho} -\frac{1}{2}u^\mu x^\nu \partial_\mu h \partial_\nu h+x^\mu u^\nu \partial_\nu h_\mu^\lambda \partial_\lambda h+u^\mu x^\nu \partial_\nu h_\mu^\lambda \partial_\lambda h\right.\right.\nonumber\\
&\left.\left.- u^\mu x^\nu \partial_\lambda h_{\mu\nu} \partial^\lambda h-u^\mu x^\nu \partial_\lambda h_\mu^\lambda \partial_\rho h_\nu^\rho - x^\mu u^\nu\partial_\nu h_{\mu\lambda}\partial_\rho h^{\lambda\rho} - u^\mu x^\nu\partial_\nu h_{\mu\lambda}\partial_\rho h^{\lambda\rho}\right.\right. \left.\left.+u^\mu x^\nu\partial_\lambda h_{\mu\nu}\partial_\rho h^{\lambda\rho}+u^\mu x^\nu\partial_\rho h_{\nu\lambda}\partial^\rho h_\mu^\lambda\right)\right.\nonumber\\
&\left.\alpha_3\left(\frac{1}{2}x^\mu x^\nu \partial_\mu h^{\lambda\rho}\partial_\nu h_{\lambda\rho} -\frac{1}{2}x^\mu x^\nu \partial_\mu h \partial_\nu h +2x^\mu x^\nu \partial_\nu h_\mu^\lambda \partial_\lambda h - x^\mu x^\nu \partial_\lambda h_{\mu\nu} \partial^\lambda h -x^\mu x^\nu \partial_\lambda h_\mu^\lambda \partial_\rho h_\nu^\rho -2 x^\mu x^\nu\partial_\nu h_{\mu\lambda}\partial_\rho h^{\lambda\rho} \right.\right.\nonumber\\
&\left.\left. +x^\mu x^\nu\partial_\lambda h_{\mu\nu}\partial_\rho h^{\lambda\rho}+x^\mu x^\nu\partial_\rho h_{\nu\lambda}\partial^\rho h_\mu^\lambda \right)\right. \left.+\alpha_4\left(\frac{1}{2}u^\mu u^\nu x^\lambda x^\rho \partial_\mu h_\lambda^\sigma \partial_\nu h_{\rho\sigma} -\frac{1}{2}u^\mu u^\nu x^\lambda x^\rho\partial_\mu h_{\lambda\rho} \partial_\nu h + \frac{1}{2}u^\mu u^\nu x^\lambda x^\rho \partial_\lambda h_\mu^\sigma\partial_\rho h_{\nu\sigma}\right.\right.\nonumber\\
&\left.\left.-u^\mu u^\nu x^\lambda x^\rho\partial_\nu h_\mu^\sigma\partial_\rho h_{\lambda\sigma} + u^\mu u^\nu x^\lambda x^\rho\partial_\nu h_{\mu\lambda}\partial_\rho h -\frac{1}{2}u^\mu u^\nu x^\lambda x^\rho\partial_\lambda h_{\mu\nu} \partial_\rho h+ u^\mu u^\nu x^\lambda x^\rho\partial_\nu h_\mu^\sigma\partial_\sigma h_{\lambda\rho} + u^\mu u^\nu x^\lambda x^\rho \partial_\rho h_{\lambda\sigma}\partial^\sigma h_{\mu\nu} \right.\right.\nonumber\\
&\left.\left. -\frac{1}{2}u^\mu u^\nu x^\lambda x^\rho \partial_\sigma h_{\lambda\rho}\partial^\sigma h_{\mu\nu}\right.\right. \left.\left.-u^\mu u^\nu x^\lambda x^\rho\partial_\nu h_{\rho\sigma}\partial^\sigma h_{\mu\lambda}-u^\mu u^\nu x^\lambda x^\rho\partial_\rho h_{\nu\sigma}\partial^\sigma h_{\mu\lambda} +\frac{1}{2}u^\mu u^\nu x^\lambda x^\rho\partial_\sigma h_{\nu\rho}\partial^\sigma h_{\mu\lambda}\right)\right],
\end{align}
\end{widetext}
where we have again redefined the perturbation $h_{\mu\nu}$ to eliminate an overall free factor in the action, and the $\alpha_n$ are arbitrary constants. There are now 3 additional parameters, the $\alpha_n$, which enter the action explicitly multiplying Lorentz-breaking terms (i.e.~those terms containing explicit factors of $u^\mu$ and $x^\mu$). As in Section \ref{SecFP}, these parameters are artefacts that arise from constructing a quadratic action with background vectors and tensors that do not respect the desired global background symmetries. One can see that the action given by eq.~(\ref{SrecoverFP}) can be recovered by imposing spatial isotropy, and thus requiring that $\alpha_2=\alpha_3=\alpha_4=0$, and recover eq.~(\ref{SFPfinal}) by imposing further invariance under boosts and hence setting $\alpha_1=0$.

From these two previous examples we conclude that it is a consistency condition to construct the quadratic action for perturbations using projectors that preserve the global background symmetries. This is because the general tensors of step 2 $\mathcal{A}$, $\mathcal{B}$, $\mathcal{C}$, etc, can only come from the background fields, and hence they must preserve the same global symmetries. 

\subsection{Axisymmetric Bianchi-I}
We can explore the role that the global symmetries of the background play further by constructing the most general diffeomorphism-invariant quadratic gravitational action in an anisotropic universe. For the background, we will consider an axisymmetric Bianchi-I model \cite{Ellis:1968vb,Ellis:1998ct,jacobs1969bianchi}, such that the line element is given by:
\begin{align}
ds^2=-dt^2+a(t)^2dx^2+b(t)^2\left(dy^2+dz^2\right).
\end{align}
Unlike the case of an isotropic universe, anisotropic universes permit dynamic vacuum solutions (e.g.~the Kasner models in GR \cite{Kasner:1921zz,nla.cat-vn2229611}); for simplicity we will consider such an anisotropic vacuum universe here. In this case, the right set of projectors to be chosen are those from a 1+1+2 split of space-time as in eq.~(\ref{bianchisplit}). The projectors $u^\mu$, $x^\mu$, and $\gamma_{\mu\nu}$ are, however, non trivial due to the dynamical nature of the background. Explicitly, they are given by:
\begin{align}
u_\mu=(-1,\textbf{0})_\mu,\\
x_\mu=(0,a,0,0)_\mu,\\
\gamma_{ij}=b^2\delta_{ij},\\
\gamma_{\mu 0}=\gamma_{\mu 1}=0,
\end{align}
where $i, j$ now run over coordinates 2 and 3 such that $u^\mu$, $x^\mu$, and $\gamma_{\mu\nu}$ are mutually orthogonal. Having chosen a set of projectors in accordance with step 1 of the method outlined throughout this paper , we can now move onto step 2 and write the most general quadratic action leading to second-order derivative equations of motion. This action can be written as:
\begin{align}
S_G^{(2)}=\int d^4x \,a\,b^2&\left[\mathcal{D}^{\mu\nu\alpha\beta}h_{\mu\nu}h_{\alpha\beta}+\mathcal{E}^{\mu\nu\alpha\beta\delta}\bar{\nabla}_\delta h_{\mu\nu} h_{\alpha\beta}\right.\nonumber\\
&\left. + \mathcal{F}^{\mu\nu\alpha\beta\kappa\delta}\bar{\nabla}_{\kappa}h_{\mu\nu}\bar{\nabla}_{\delta}h_{\alpha\beta}\right],\label{Sbianchi}
\end{align}
where the tensors $\mathcal{D}$, $\mathcal{E}$, and $\mathcal{F}$ are given in Appendix \ref{appendixBianchi}. The action given by eq.~(\ref{Sbianchi}) depends on 122 free functions of time, a large increase from the 26 free functions that were needed in Section \ref{secGR}. We additionally have 2 free functions of time from the background - the scale factors $a$ and $b$.

Having obtained our most general quadratic action for the anisotropic Bianchi-I background, we can now proceed to step 3 and impose diffeomorphism invariance on the action (\ref{Sbianchi}), with the metric transforming as in eq.~(\ref{hgaugetransformation}). As in the previous sections, we obtain a number of Noether constraints that reduce the number of free parameters present in our theory from 124 (122 from eq.~(\ref{Sbianchi}) and 2 from the background) to one free function of time $M^2$ and two constants $c_1$ and $c_2$ (a full list of the Noether constraints can be found in Appendix \ref{appendixBianchi}), whilst a relation between the two scale factors $a$ and $b$ is also found. The final gauge invariant action is given by:
\begin{widetext}
\begin{align}
S_G^{(2)}=\int d^4x\,a\,b^2\,M^2&\left[-\frac{1}{2} h_{\mu\nu} h^{\mu\nu} H_2^2+\frac{1}{4} h ^2 H_2^2-\frac{1}{2} h^{\nu\lambda} u^\mu
 \bar{\nabla}_\lambda h_{\mu\nu} H_2+\left(\frac{4 c_1 }{M^2}-\frac{1}{4}\right) h u^\mu x^\nu
 x^\lambda \bar{\nabla}_\lambda h_{\mu\nu} H_2+\frac{1}{4} h u^\mu \bar{\nabla}_\lambda h_\mu^\lambda
 H_2\right.\nonumber\\
 &+\frac{1}{2} h_\nu^\sigma u^\mu x^\nu x^\lambda \bar{\nabla}_\lambda h_{\mu\sigma} H_2-\frac{1}{2}
 h_\mu^\sigma u^\mu u^\nu u^\lambda \bar{\nabla}_\lambda h_{\nu\sigma} H_2+\frac{4 c_1 h_\mu^\sigma u^\mu
 x^\nu x^\lambda \bar{\nabla}_\lambda h_{\nu\sigma} H_2}{M^2}\nonumber\\
 &+\left(\frac{1}{2}-\frac{2 c_1 }{M^2}\right)
 h_{\mu\sigma} u^\mu u^\nu u^\lambda x^\sigma x^\alpha \bar{\nabla}_\lambda h_{\nu\alpha} 
 H_2+\left(\frac{1}{2}-\frac{2 c_1 }{M^2}\right) h_\nu^\sigma u^\mu x^\nu x^\lambda
 \bar{\nabla}_\sigma h_{\mu\lambda} H_2-\frac{1}{4} h_{\nu\lambda} u^\mu x^\nu x^\lambda
 \bar{\nabla}_\sigma h_\mu^\sigma H_2\nonumber\\
 &+\left(\frac{ c_1 }{M^2}-\frac{1}{2}\right) h_\mu^\sigma u^\mu u^\nu u^\lambda
 \bar{\nabla}_\sigma h_{\nu\lambda} H_2+\frac{c h_\mu^\sigma u^\mu x^\nu x^\lambda
 \bar{\nabla}_\sigma h_{\nu\lambda} H_2}{M^2}+\frac{1}{4} h_{\mu\nu} u^\mu u^\nu u^\lambda
 \bar{\nabla}_\sigma h_\lambda^\sigma H_2\nonumber\\
 &+\left(-\frac{2 c_1 }{M^2}-\frac{1}{4}\right) h_{\nu\lambda} u^\mu x^\nu
 x^\lambda x^\sigma x^\alpha \bar{\nabla}_\alpha h_{\mu\sigma} H_2+\left(\frac{1}{2}-\frac{ c_1 }{M^2}\right)
 h_{\mu\sigma} u^\mu u^\nu u^\lambda x^\sigma x^\alpha \bar{\nabla}_\alpha h_{\nu\lambda} H_2\nonumber\\
 &+\left(\frac{12
 c_1 }{M^2}-\frac{1}{4}\right) h_{\mu\nu} u^\mu u^\nu u^\lambda x^\sigma x^\alpha
 \bar{\nabla}_\alpha h_{\lambda\sigma} H_2-\frac{5 c_1 h_{\mu\nu} u^\mu x^\nu x^\lambda x^\sigma x^\alpha
 \bar{\nabla}_\alpha h_{\lambda\sigma} H_2}{M^2}\nonumber\\
 &+\left(\left(\frac{ c_1 }{M^2}-\frac{3}{2}\right) H_2^2+\frac{2 c_2
 c_1 }{M^2}\right) h_\mu^\lambda h_{\nu\lambda} u^\mu u^\nu+\frac{\left(H_2^2 \left(M^2-2 c_1 \right)-4 c_2
 c_1 \right) h_{\mu\nu} h u^\mu u^\nu}{2 M^2}\nonumber\\
 &+\left(\frac{2 c_2 c_1 }{M^2}-\frac{3 H_2^2}{4}\right)
 h_{\mu\nu} h_{\lambda\sigma} u^\mu u^\nu u^\lambda u^\sigma +\frac{\left(\left(M^2-3 c_1 \right) H_2^2+2 c_2
 c_1 \right) h_\mu^\lambda h_{\nu\lambda} x^\mu x^\nu}{M^2}\nonumber\\
 &+\frac{\left(4 c_2 c_1 -H_2^2 \left(M^2+2
 c_1 \right)\right) h_{\mu\nu} h x^\mu x^\nu}{2 M^2}+\left(\left(\frac{3}{2}-\frac{8 c_1 }{M^2}\right)
 H_2^2+\frac{4 c_2 c_1 }{M^2}\right) h_{\mu\lambda} h_{\nu\sigma} u^\mu u^\nu x^\lambda x^\sigma \nonumber\\
 &+\left(\frac{8 c_2
 c_1 }{M^2}-\frac{H_2^2}{2}\right) h_{\mu\nu} h_{\lambda\sigma} u^\mu u^\nu x^\lambda x^\sigma +\left(H_2^2
 \left(\frac{4 c_1 }{M^2}-\frac{1}{4}\right)-\frac{4 c_2 c_1 }{M^2}\right) h_{\mu\nu} h_{\lambda\sigma} x^\mu x^\nu
 x^\lambda x^\sigma \nonumber\\
 &+\left(-\frac{ c_1 }{2 M^2}-\frac{1}{8}\right) u^\mu u^\nu \bar{\nabla}_\mu h^{\lambda\sigma} 
 \bar{\nabla}_\nu h_{\lambda\sigma} +\left(\frac{ c_1 }{2 M^2}+\frac{1}{8}\right) x^\mu x^\nu
 \bar{\nabla}_\mu h^{\lambda\sigma} \bar{\nabla}_\nu h_{\lambda\sigma}\nonumber\\
 &+\left(\frac{ c_1 }{M^2}
 +\frac{1}{4}\right) u^\mu u^\nu
 x^\lambda x^\sigma \bar{\nabla}_\mu h_\lambda^\alpha \bar{\nabla}_\nu h_{\sigma\alpha} +\left(\frac{ c_1 }{2
 M^2}+\frac{1}{8}\right) u^\mu u^\nu \bar{\nabla}_\mu h 
 \bar{\nabla}_\nu h\nonumber\\
 &+\left(-\frac{ c_1 }{2 M^2}-\frac{1}{8}\right) x^\mu x^\nu
 \bar{\nabla}_\mu h \bar{\nabla}_\nu h +\left(-\frac{ c_1 }{M^2}-\frac{1}{4}\right) u^\mu
 u^\nu x^\lambda x^\sigma \bar{\nabla}_\mu h_{\lambda\sigma} \bar{\nabla}_\nu h +\frac{1}{8}
 \bar{\nabla}_\nu h \bar{\nabla}^\nu h\nonumber\\
 &+\left(H_2 \left(\frac{1}{4}-\frac{ c_1 }{2 M^2}\right)-\frac{c_2
 c_1 }{H_2 M^2}\right) h u^\mu u^\nu u^\lambda \bar{\nabla}_\lambda h_{\mu\nu}+\left(\frac{c_2
 c_1 }{H_2 M^2}+H_2 \left(\frac{5 c_1 }{2 M^2}-\frac{1}{4}\right)\right) h_{\sigma\alpha} u^\mu u^\nu u^\lambda
 x^\sigma x^\alpha \bar{\nabla}_\lambda h_{\mu\nu}\nonumber\\
 &-\frac{1}{4} u^\mu u^\nu \bar{\nabla}_\nu h 
 \bar{\nabla}_\lambda h_\mu^\lambda+\frac{1}{4} x^\mu x^\nu \bar{\nabla}_\nu h 
 \bar{\nabla}_\lambda h_\mu^\lambda+\frac{1}{4} \bar{\nabla}_\mu h^{\mu\nu} \bar{\nabla}_\lambda h_\nu^\lambda -\frac{1}{4}
 \bar{\nabla}^\nu h \bar{\nabla}_\lambda h_\nu^\lambda +\left(-\frac{2 c_1 }{M^2}-\frac{1}{4}\right) u^\mu u^\nu
 \bar{\nabla}_\nu h_\mu^\lambda \bar{\nabla}_\lambda h\nonumber\\
 &+\left(\frac{2 c_1 }{M^2}+\frac{1}{4}\right) x^\mu 
 x^\nu \bar{\nabla}_\nu h_\mu^\lambda \bar{\nabla}_\lambda h +\left(\frac{ c_1 }{M^2}+\frac{1}{4}\right) u^\mu
 u^\nu \bar{\nabla}_\lambda h \bar{\nabla}^\lambda h_{\mu\nu}+\left(-\frac{ c_1 }{M^2}-\frac{1}{4}\right)
 x^\mu x^\nu \bar{\nabla}_\lambda h \bar{\nabla}^\lambda h_{\mu\nu}\nonumber\\
 &-\frac{1}{8}
 \bar{\nabla}_\lambda h_{\mu\nu} \bar{\nabla}^\lambda h^{\mu\nu} +\left(\frac{ c_1 }{M^2}-\frac{1}{4}\right) u^\mu u^\nu
 x^\lambda x^\sigma x^\alpha x^\beta \bar{\nabla}_\nu h_{\alpha\beta} 
 \bar{\nabla}_\sigma h_{\mu\lambda}+\left(\frac{2 c_1 }{M^2}+\frac{1}{4}\right) u^\mu u^\nu x^\lambda x^\sigma 
 \bar{\nabla}_\nu h \bar{\nabla}_\sigma h_{\mu\lambda}\nonumber\\
 & +\frac{1}{4} u^\mu u^\nu
 \bar{\nabla}_\lambda h_\mu^\lambda \bar{\nabla}_\sigma h_\nu^\sigma -\frac{1}{4} x^\mu x^\nu
 \bar{\nabla}_\lambda h_\mu^\lambda \bar{\nabla}_\sigma h_\nu^\sigma +\left(\frac{ c_1 }{M^2}+\frac{1}{4}\right) u^\mu u^\nu
 x^\lambda x^\sigma \bar{\nabla}_\lambda h_\mu^\alpha \bar{\nabla}_\sigma h_{\nu\alpha} \nonumber\\
 &+\left(\frac{2
 c_1 }{M^2}+\frac{1}{2}\right) u^\mu u^\nu \bar{\nabla}_\nu h_\mu^\lambda
 \bar{\nabla}_\sigma h_\lambda^\sigma +\left(-\frac{2 c_1 }{M^2}-\frac{1}{2}\right) x^\mu x^\nu
 \bar{\nabla}_\nu h_\mu^\lambda \bar{\nabla}_\sigma h_\lambda^\sigma +\left(-\frac{ c_1 }{M^2}-\frac{1}{4}\right) u^\mu
 u^\nu \bar{\nabla}^\lambda h_{\mu\nu} \bar{\nabla}_\sigma h_\lambda^\sigma \nonumber\\
 &+\left(\frac{ c_1 }{M^2}+\frac{1}{4}\right) x^\mu 
 x^\nu \bar{\nabla}^\lambda h_{\mu\nu} \bar{\nabla}_\sigma h_\lambda^\sigma +\left(-\frac{2 c_1 }{M^2}-\frac{1}{2}\right)
 u^\mu u^\nu x^\lambda x^\sigma \bar{\nabla}_\nu h_\mu^\alpha \bar{\nabla}_\sigma h_{\lambda\alpha} +\frac{1}{4} u^\mu
 u^\nu x^\lambda x^\sigma \bar{\nabla}_\nu h_{\mu\lambda} 
 \bar{\nabla}_\sigma h \nonumber\\
 &+\left(-\frac{ c_1 }{M^2}-\frac{1}{4}\right) u^\mu u^\nu x^\lambda x^\sigma 
 \bar{\nabla}_\lambda h_{\mu\nu} \bar{\nabla}_\sigma h +\frac{ c_1u^\mu u^\nu
 \bar{\nabla}_\lambda h_{\nu\sigma} \bar{\nabla}^\sigma h_\mu^\lambda}{M^2}-\frac{ c_1 x^\mu x^\nu
  \bar{\nabla}_\lambda h_{\nu\sigma}  \bar{\nabla}^\sigma h_\mu^\lambda}{M^2}\nonumber\\
  &+\left(-\frac{ c_1 }{M^2}-\frac{1}{4}\right) u^\mu
  u^\nu \bar{\nabla}_\sigma h_{\nu\lambda}  \bar{\nabla}^\sigma h_\mu^\lambda+\left(\frac{ c_1 }{M^2}+\frac{1}{4}\right) x^\mu 
  x^\nu \bar{\nabla}_\sigma h_{\nu\lambda}  \bar{\nabla}^\sigma h_\mu^\lambda\nonumber\\
  &+\left(\frac{1}{4}-\frac{3 c_1 }{M^2}\right)
  u^\mu u^\nu u^\lambda u^\sigma  x^\alpha x^\beta  \bar{\nabla}_\sigma h_{\lambda\beta} 
  \bar{\nabla}_\alpha h_{\mu\nu}+\left(\frac{1}{4}-\frac{ c_1 }{M^2}\right) u^\mu u^\nu x^\lambda x^\sigma 
  \bar{\nabla}_\nu h_{\lambda\sigma}  \bar{\nabla}_\alpha h_\mu^\alpha\nonumber\\
  &-\frac{1}{2} u^\mu u^\nu x^\lambda x^\sigma 
  \bar{\nabla}_\sigma h_{\mu\lambda}  \bar{\nabla}_\alpha h_\nu^\alpha +\left(H_2 \left(\frac{ c_1 }{2
  M^2}-\frac{1}{4}\right)-\frac{c_1 c_2 }{H_2 M^2}\right) h_{\mu\nu} u^\mu u^\nu u^\lambda u^\sigma  u^\alpha 
  \bar{\nabla}_\alpha h_{\lambda\sigma} \nonumber\\
  &+\left(\frac{3 c_1 }{M^2}+\frac{1}{4}\right) u^\mu u^\nu x^\lambda x^\sigma 
  \bar{\nabla}_\nu h_\mu^\alpha \bar{\nabla}_\alpha h_{\lambda\sigma}-\frac{1}{2} u^\mu u^\nu x^\lambda x^\sigma 
  \bar{\nabla}_\nu h_{\mu\lambda}  \bar{\nabla}_\alpha h_\sigma^\alpha +\left(\frac{ c_1 }{M^2}+\frac{1}{4}\right) u^\mu
  u^\nu u^\lambda u^\sigma \bar{\nabla}_\lambda h_{\mu\nu}
  \bar{\nabla}_\alpha h_\sigma^\alpha \nonumber\\
  &+\left(\frac{1}{4}-\frac{ c_1 }{M^2}\right) u^\mu u^\nu x^\lambda x^\sigma 
  \bar{\nabla}_\lambda h_{\mu\nu} \bar{\nabla}_\alpha h_\sigma^\alpha +\left(\frac{ c_1 }{M^2}+\frac{1}{4}\right) x^\mu 
  x^\nu x^\lambda x^\sigma \bar{\nabla}_\lambda h_{\mu\nu}
  \bar{\nabla}_\alpha h_\sigma^\alpha \nonumber\\
  &+\left(-\frac{ c_1 }{M^2}-\frac{1}{4}\right) u^\mu u^\nu u^\lambda u^\sigma 
  \bar{\nabla}_\sigma h_{\lambda\alpha}  \bar{\nabla} h_{\mu\nu}+\left(\frac{3 c_1 }{M^2}+\frac{1}{4}\right) u^\mu
  u^\nu x^\lambda x^\sigma  \bar{\nabla}_\sigma h_{\lambda\alpha} 
  \bar{\nabla}^\alpha h_{\mu\nu}\nonumber\\
  &+\left(-\frac{ c_1 }{M^2}-\frac{1}{4}\right) x^\mu x^\nu x^\lambda x^\sigma 
  \bar{\nabla}_\sigma h_{\lambda\alpha}  \bar{\nabla}^\alpha h_{\mu\nu}+\left(-\frac{ c_1 }{M^2}-\frac{1}{4}\right) u^\mu
  u^\nu x^\lambda x^\sigma  \bar{\nabla}_\alpha h_{\lambda\sigma}  \bar{\nabla}^\alpha h_{\mu\nu}\nonumber\\
  &-\frac{2 c_1 u^\mu
  u^\nu x^\lambda x^\sigma  \bar{\nabla}_\nu h_{\sigma\alpha}  \bar{\nabla}^\alpha h_{\mu\lambda} }{M^2}-\frac{2 c_1 
  u^\mu u^\nu x^\lambda x^\sigma  \bar{\nabla}_\sigma h_{\nu\alpha} 
  \bar{\nabla}^\alpha h_{\mu\lambda} }{M^2}+\left(\frac{ c_1 }{M^2}+\frac{1}{4}\right) u^\mu u^\nu x^\lambda x^\sigma 
  \bar{\nabla}_\alpha h_{\nu\sigma}  \bar{\nabla}^\alpha h_{\mu\lambda} \nonumber\\
  &+\left(\frac{1}{4}-\frac{ c_1 }{M^2}\right) u^\mu
  u^\nu x^\lambda x^\sigma  x^\alpha x^\beta  \bar{\nabla}_\nu h_{\mu\lambda} 
  \bar{\nabla}_\beta h_{\sigma\alpha} +\left(\frac{3 c_1}{M^2}-\frac{1}{4}\right) u^\mu u^\nu u^\lambda u^\sigma 
  x^\alpha x^\beta \bar{\nabla}_\lambda h_{\mu\nu} \bar{\nabla}_\beta h_{\sigma\alpha}\nonumber\\
  & -\frac{ c_1\left(H_2^2+2
  c_2\right) h u^\mu x^\nu x^\lambda \bar{\nabla}_\mu h_{\nu\lambda} }{2 M^2 H_2}+\frac{c_1
  \left(H_2^2+2 c_2\right) h_{\nu\lambda} u^\mu x^\nu x^\lambda x^\sigma  x^\alpha
  \bar{\nabla}_\mu h_{\sigma\alpha} }{2 M^2 H_2}\nonumber\\
  &\left.+\frac{ c_1\left(H_2^2-2 c_2\right) h_{\mu\nu} u^\mu u^\nu u^\lambda
  x^\sigma  x^\alpha \bar{\nabla}_\lambda h_{\sigma\alpha} }{2 M^2 H_2}\right].
\end{align} 
\end{widetext}

$M^2$ and $c_1$ arise from the 122 free parameters in the action (\ref{Sbianchi}), whilst $c_2$ is an integration constant which comes from the relation found between the two scale factors:
\begin{align}
H_1=\frac{c_2}{H_2}-\frac{1}{2}H_2,
\end{align}
where we have defined $H_1=\frac{\dot{a}}{a}$ and $H_2=\frac{\dot{b}}{b}$. By setting $c_2=0$ we find that $H_1=-\frac{1}{2}H_2$. This relation corresponds to the non-trivial Kasner solution for an axisymmetric vacuum universe in General Relativity \cite{Kasner:1921zz,nla.cat-vn2229611, Centrella:1986tf, Petersen:2015mva}:
\begin{align}
ds^2=-dt^2+t^{-\frac{2}{3}}dx^2+t^{\frac{4}{3}}\left(dy^2+dz^2\right).\label{kasnermetric}
\end{align}
We can further set $M^2=-4c_1=M_{Pl}^2$ and, after making some integrations by parts on the remaining terms, recover the exact Kasner solution (i.e.~the quadratic expansion of the Einstein-Hilbert action about the background given by eq.~(\ref{kasnermetric})):
\begin{widetext}
\begin{align}
S_{GR}^{(2)}=\int d^4x \,t\,M^2&\left[-\frac{1}{9t^2}h_{\mu\nu}h^{\mu\nu}+\frac{1}{9t^2}h^2-\frac{1}{3t^2}h_\mu^\alpha h_{\nu\alpha}u^\mu u^\nu+\frac{1}{3t^2}h_\mu^\alpha h_{\nu\alpha}x^\mu x^\nu+\frac{1}{3t^2}hh_{\mu\nu}u^\mu u^\nu-\frac{1}{3t^2}hh_{\mu\nu}x^\mu x^\nu\right.\nonumber\\
& +\frac{2}{3t^2}h_{\mu\nu}h_{\alpha\beta}u^\mu u^\alpha x^\nu x^\beta -\frac{2}{3t^2}h_{\mu\nu}h_{\alpha\beta}u^\mu u^\nu x^\alpha x^\beta+\frac{1}{8}\bar{\nabla}_\mu h \bar{\nabla}^\mu h +\frac{1}{4}\bar{\nabla}_\mu h^{\mu\alpha}\bar{\nabla}^\nu h_{\nu\alpha}-\frac{1}{4}\bar{\nabla}^\mu h \bar{\nabla}^\nu h_{\mu\nu}\nonumber\\
&\left.-\frac{1}{8}\bar{\nabla}_\alpha h_{\mu\nu} \bar{\nabla}^\alpha h^{\mu\nu}\right].
\end{align}
\end{widetext}
Thus, as in Section \ref{secGR}, the correct general relativistic solution can be found by a specific choice for the remaining free parameters in our theory. Unlike the resulting gauge invariant action found in \ref{secGR}, however, the action found for an axisymmetric Bianchi-I vacuum universe depends not only on a free function of time $M^2$ (as well as the background scale factor $a$), but also on two constants $c_1$ and $c_2$. This increased number of parameters in the final theory is a result of the reduced symmetry of our background space-time (from a homogenous and isotropic FLRW space-time to an anisotropic axisymmetric Bianchi-I space-time).

As we have seen in the previous examples, the symmetry of the background plays a crucial role on determining the final number of relevant free parameters in the quadratic action for perturbation. In general, the less symmetric the background, the more free parameters we will get (or at least the same number). As we have seen, we impose a given background symmetry by choosing the appropriate basis of background vectors and tensors that respect the symmetry, and construct the most general action in step 2 using that basis. Doing this is crucial for consistency as the coefficients of this general action (such as $\mathcal{A}^{\mu\nu\alpha\beta\gamma\delta}$ and $\mathcal{B}^{\alpha\beta\gamma\delta}$) can only come from the background fields and their derivatives, and hence they must respect the same symmetries. 


\section{Conclusion}\label{conclusion}
In this paper we have presented a covariant approach for constructing quadratic actions for linear perturbations, for a given set of fields, background global symmetries, and gauge local symmetries. We have discussed the relevance in distinguishing gauge and global symmetries and the role they play in the final construction of quadratic actions of perturbations. 

The approach presented in this paper is divided in 3 steps. In step 1 we choose the background on which perturbations propagate. This background will usually have a certain set of global symmetries, i.e. rigid symmetries that do not depend on space and time. For instance, if the background is Minkowski the global symmetries will be given by the Poincare group, but if the background is FLRW, the symmetries are spatial rotations and translations (isotropy and homogeneity, respectively). In step 2 we construct the most general quadratic action for perturbations that lead to a chosen maximum number of derivatives in the equations of motion. This general action will have free coefficient multiplying the different possible quadratic interaction terms of the perturbation fields. These coefficients come from the background fields and their derivatives, and hence they must satisfy the same global symmetries of the background in order to be consistent. Therefore, the background symmetries play a crucial role in step 2. We achieve this consistency by choosing an appropriate basis of background projectors to use to write the general coefficients of the general quadratic action. Finally, in step 3, we impose that the general action of step 2 is invariant under certain local gauge transformations and, hence, in this step gauge symmetries are the ones playing a crucial role. We impose gauge symmetries by finding the relevant set of Noether identities associated to the symmetry and enforcing that they vanish. This leads to a set of relations between the free coefficients of the quadratic action in such a way that the final action is invariant under the desired local symmetries.   

The covariant action approach presented in this paper is general and systematic and we have shown how it can be applied to cosmology to construct general parametrised actions linear perturbations for different families of modified gravity models: scalar-tensor and vector-tensor diffeomorphism invariant theories. Since we have imposed gauge invariance on the covariant set of perturbation fields, we have hence made scalar and vector perturbations gauge invariant, and we have presented their corresponding actions in this paper. In the case of scalar-tensor theories, we have recovered the same well-known result of previous works, but for vector-tensor theories we have extended the results of \cite{Lagos:2016wyv} and found that the action for scalar perturbations depends on 9 free parameters instead on 10, once gauge invariance on vector perturbations is imposed. We have also shown explicitly the action for vector and tensor perturbations, which are found to depend only on 5 and 1 free parameters, respectively. These results highlight the fact that scalar perturbations are essential for constraining modified gravity as they are the ones containing the most information on the free parameters, but the search of signatures in tensor or vector modes could be used complementary to improve constraints on some of the free parameters. 

The power of our method is that it can be applied to any type of background, with the example of an axisymmetric Bianchi-I vacuum model given explicitly. In particular, it can be applied to non-cosmological backgrounds such a black hole space-time of various forms and guises \cite{Ripley:2017kqg}, linking cosmological tests to tests on astrophysical scales (see \cite{Cornish:2011ys,Loutrel:2014vja,Sampson:2013wia} for attempts at connecting different regimes). In that case it should be possible to determine the most general set of linear perturbations for a given field content which will play a role in ringdown \cite{Berti:2009kk}; this approach could generalise the usual quasinormal analysis of general relativistic black holes \cite{Kobayashi:2012kh,Kobayashi:2014wsa} and extend the analysis of gravitational wave experiments in the case of black hole mergers. In particular, it could give us a general method for exploring violations of the no-hair theorem in extended theories of gravity \cite{Dreyer:2003bv,Cardoso:2016ryw}. Furthermore, the use of a fully covariant approach lends great transparency to the resulting gauge invariant actions calculated using the method discussed in this paper. For example, it is clear which terms in the final action originate due to broken symmetries of the background (rather than being contracted with the full metric). This allow us, for instance, to easily recover the action for backgrounds with more symmetries from actions for backgrounds with less symmetries.



\textit{Acknowledgements ---} We thank T. Baker, E. Bellini and J. Noller for useful conversations. OJT is grateful to J. Bonifacio for his help in using the $x$\textit{Tras} package for Mathematica \cite{Nutma:2013zea}, which was used in the computation of some of the results presented here. OJT was supported by the Science and Technology Facilities Council (STFC) Project Reference No. 1804725. ML was funded by Becas Chile, CONICYT. PGF acknowledges support from Leverhulme, STFC, BIPAC and the ERC. 

\appendix
\section{Noether constraints for scalar-tensor theories}

\subsection{Covariant quantities}\label{appendixS1}
With the introduction of a matter sector, the background space-time will no longer be flat. Thus we need expressions for the Christoffel symbols and curvature tensors of the background \textit{in terms of the background quantities} to properly evaluate the Noether constraints arising from the variation of (\ref{Sgencosmology}). The relevant expressions can be shown to be:
\begin{align}
\Gamma^\rho_{\,\mu\nu}=&H(\gamma_{\mu\nu}u^\rho-\gamma^\rho_{\,\mu}u_\nu-\gamma^\rho_{\,\nu}u_\mu)\\
\bar{\nabla}_\mu u_\nu=&H\gamma_{\mu\nu}\\
\bar{\nabla}_\mu\gamma_{\alpha\beta}=&u_\alpha\bar{\nabla}_\mu u_\beta + u_\beta\bar{\nabla}_\mu u_\alpha\\
\bar{R}^\rho_{\,\sigma\mu\nu}=&\dot{H}(-u_\mu u^\rho \gamma_{\nu\sigma}+u_\nu u^\rho \gamma_{\mu\sigma}+\gamma^\rho_\nu u_\mu u_\sigma-\gamma^\rho_\mu u_\nu u_\sigma)\nonumber\\
&+H^2(u_\nu u^\rho \gamma_{\mu\sigma}-u_\mu u^\rho \gamma_{\nu\sigma}-\gamma^\rho_\mu u_\nu u_\sigma+\gamma^\rho_\nu u_\mu u_\sigma\nonumber\\
& +\gamma^\rho_\mu\gamma_{\sigma\nu}-\gamma^\rho_\nu\gamma_{\sigma\mu})\\
\bar{R}_{\mu\nu}=&-3\left(\dot{H}+H^2\right)u_\mu u_\nu+\left(3H^2+\dot{H}\right)\gamma_{\mu\nu}\\
\bar{R}=&12H^2+6\dot{H}.
\end{align}
$H=\frac{d\log{a}}{dt}$ is the Hubble parameter, $\bar{R}^\rho_{\,\sigma\mu\nu}$ is the Riemann curvature tensor, $\bar{R}_{\mu\nu}=\bar{R}^\rho_{\mu\rho\nu}$ is the Ricci tensor, and $\bar{R}=\bar{g}^{\mu\nu}\bar{R}_{\mu\nu}$ is the Ricci scalar. Note that, unlike in the 1+3 covariant formalism introduced in \cite{Tsagas:2007yx}, we have not introduced `shear' or `velocity' tensors, nor a `volume expansion' scalar or `acceleration' vector. Every background tensor can be expressed in terms of functions of time and the projectors $u^\mu$ and $\gamma_{\mu\nu}$. These expressions are, however, only valid in the chosen coordinate basis. 

\subsection{Solutions}\label{appendixS2}
The following Noether constraints are obtained in Section \ref{secST} for the $A_n$, $B_n$, and $C_n$:
\begin{alignat}{3}
&A_1&&=-\frac{\dot{\bar{\varphi}}^2}{16}-H\dot{\bar{\chi}} D_{\chi5}\nonumber\\
&A_2&&=\frac{1}{8}\left(\dot{\bar{\varphi}}^2+8H\dot{\bar{\chi}}D_{\chi5}-32H^2C_1\right)\nonumber\\
&A_3&&=\frac{1}{8}\left(-\dot{\bar{\varphi}}^2-32H^2C_1-32H^2C_5-8H\dot{\bar{\chi}} D_{\chi2} - 16H\dot{\bar{\chi}} D_{\chi5}\right)\nonumber\\
&A_4&&=\frac{1}{4}\left(\dot{\bar{\varphi}}^2-32H^2C_1-16\dot{H}C_5+4H\dot{\bar{\chi}}D_{\chi5}\right)\nonumber\\
&A_5&&=\frac{1}{16}\left(-\dot{\bar{\varphi}}^2-96H^2C_5+4C_{\chi1}\dot{\bar{\chi}}^2-24H\dot{\bar{\chi}}D_{\chi2}\right)\nonumber\\
&B_1&&=\dot{\bar{\chi}}D_{\chi5}-4HC_1\nonumber\\
&B_3&&=-\dot{\bar{\chi}}D_{\chi5}\nonumber\\
&B_3&&=4HC_5+\frac{1}{2}\dot{\bar{\chi}}D_{\chi2}\nonumber\\
&B_4&&=-4HC_5-\dot{\bar{\chi}}D_{\chi2}\nonumber\\
&C_1&&=-H^{-1}\left(HC_5+\dot{C}_6+\frac{1}{4}\dot{\bar{\chi}}D_{\chi5}\right)\nonumber\\
&C_2&&=-C_1\nonumber\\
&C_3&&=-C_4=-2C_1\nonumber\\
&C_6&&=-C_5\nonumber\\
&C_8&&=-C_8=2C_5\nonumber\\
&C_9&&=C_{10}=0\nonumber\\
&C_{11}&&=-C_{12}=-2C_5\nonumber\\
&C_{13}&&=-C_{14}=-4C_5\nonumber\\
&C_{15}&&=C_{16}=C_{17}=0.
\end{alignat}
For the $A_{\chi\,n}$, $B_{\chi\,n}$, $C_{\chi\,n}$, and $D_{\chi\,n}$:
\begin{align}
  A_{\chi0}=&\frac{1}{\dot{\bar{\chi}}}\left(C_{\chi1}\dddot{\bar{\chi}}-3\dot{H}\dot{\bar{\chi}}C_{\chi2}+\dot{C}_{\chi1}\ddot{\bar{\chi}} + 3HC_{\chi1}\ddot{\bar{\chi}} \right.\nonumber \\ &\left.+ 6H\dot{H}D_{\chi2}+3\ddot{H}D_{\chi2}-6H\dot{H}D_{\chi5}\right)\nonumber\\
  A_{\chi1}=&3H\dot{D}_{\chi2}+3H\dot{\bar{\chi}} C_{\chi2}-\ddot{\bar{\chi}}C_{\chi1}+3H^2D_{\chi2} \nonumber \\ &-3\dot{H}D_{\chi2}+6H^2D_{\chi5}\nonumber\\
  A_{\chi2}=&-\ddot{D}_{\chi2}-4H\dot{D}_{\chi2}-2H\dot{D}_{\chi5}-\dot{\bar{\chi}}\dot{C}_{\chi2} \nonumber \\ &-3H\dot{\bar{\chi}}C_{\chi2}-\ddot{\bar{\chi}}C_{\chi2}-3H^2D_{\chi}2-\dot{H}D_{\chi2}-6H^2D_{\chi5}\nonumber\\
  &-2\dot{H}D_{\chi5}\nonumber\\
  B_{\chi1}=&C_{\chi1}\dot{\bar{\chi}} + 3HD_{\chi2}\nonumber\\
  B_{\chi2}=&-\dot{D}_{\chi2}-C_{\chi2}\dot{\bar{\chi}}-HD_{\chi2}-2HD_{\chi5}\nonumber\\
  B_{\chi3}=&2C_{\chi2}\dot{\bar{\chi}}\nonumber\\
  D_{\chi1}=&0\nonumber\\
  D_{\chi3}=&D_{\chi{2}}\nonumber\\
  D_{\chi4}=&-2D_{\chi2}\nonumber\\
  D_{\chi6}=&-D_{\chi5}.
\end{align}

In addition, a Friedmann-like equation analogous to (\ref{fried1}) is found:
\begin{align}
 -\dot{\bar{\chi}}^2C_{\chi2}=&-\frac{1}{2}\dot{\bar{\varphi}}^2+8\dot{H}C_5+HD_{\chi2} \dot{\bar{\chi}}+\dot{D_{\chi2}}\dot{\bar{\chi}}\nonumber \\ & +D_{\chi2}\ddot{\bar{\chi}}+2H\dot{\bar{\chi}}D_{\chi5}\label{friedST}.
\end{align}

Before imposing diffeomorphism invariance, our action contained 42 unknown free functions of time: the 26 $A_n$, $B_n$, and $C_n$; the 14 $A_{\chi\,n}$, $B_{\chi\,n}$, $C_{\chi\,n}$, and $D_{\chi\,n}$; the scale factor $a$, and the background value of the scalar field $\chi_0$ ($\bar{\varphi}$ is related to $a$ through (\ref{phiEOM})). 36 Noether constraints are obtained, thus leaving us with 6 unknown free functions of time in the final gauge invariant action: $C_5$, $C_{\chi1}$, $D_{\chi2}$, $D_{\chi5}$, the scale factor $a$, and the background value of the scalar field $\bar{\chi}$. We can make the following re-definitions of some of our remaining free functions to match the $\alpha_i$ described in \cite{Gleyzes:2014rba} and \cite{Bellini:2014fua}:
\begin{align}
  M^2=&-8C_5\nonumber\\
  \alpha_M=&\frac{d\log{M^2}}{d\log{a}}\nonumber\\
  \alpha_B=&-\frac{D_{\chi2}\dot{\bar{\chi}}}{HM^2}\nonumber\\
  \alpha_K=&\frac{2C_{\chi1}\dot{\bar{\chi}}^2}{H^2M^2}\nonumber\\
  \alpha_T=&-\frac{2\dot{\bar{\chi}}D_{\chi5}}{HM^2}+\alpha_M.\label{STparameters}
\end{align}
The $\alpha_i$ can be understood through the physical effects they parameterize\cite{Bellini:2014fua}.

\section{Vector-tensor gravity}

\subsection{Noether constraints}\label{appendixVT}

The following Noether constraints are obtained in Section \ref{secVT} for the $A_n$, $B_n$, and $C_n$:
\begin{widetext}
\begin{align}
A_1=&\frac{1}{16}\left(-\dot{\bar{\varphi}}^2-16HD_{\zeta7}(\dot{\bar{\zeta}}-H\bar{\zeta})\right)\nonumber\\
A_2=&\frac{1}{8}\left(\dot{\bar{\varphi}}^2-8HD_{\zeta7}(\dot{\bar{\zeta}}-H\bar{\zeta})-32H^2C_1\right)\nonumber\\
A_3=&\frac{1}{8}\left(-\dot{\bar{\varphi}}^2-32H^2C_1-32H^2C_5-4\dot{H}D_{\zeta4}\bar{\zeta} - 24H^2D_{\zeta7}\bar{\zeta} - 8H^2D_{\zeta9}\bar{\zeta} - 8H^2C_{\zeta2}\bar{\zeta}^2\right.\nonumber\\
&\left.+8H^2C_{\zeta3}\bar{\zeta}^2
+8\dot{H}C_{\zeta3}\bar{\zeta}^2-8H^2C_{\zeta5}\bar{\zeta}^2+8H^2C_{\zeta6}\bar{\zeta}^2 - 4HD_{\zeta4}\dot{\bar{\zeta}} +16HD_{\zeta7}\dot{\bar{\zeta}}+8HC_{\zeta2}\bar{\zeta}\dot{\bar{\zeta}}\right)\nonumber\\
A_4=&\frac{1}{4} \left(-\dot{\bar{\varphi}}^2+4 C_{\zeta2} \dot{\bar{\zeta}}^2-4 C_{\zeta2} \bar{\zeta}^2 H^2-12 C_{\zeta3}
  \bar{\zeta}^2 H^2-4 \dot{C_{\zeta3}} \bar{\zeta}^2 H-4 C_{\zeta3} \bar{\zeta}^2 \dot{H}-8 C_{\zeta3}
  \bar{\zeta} \dot{\bar{\zeta}} H+\right.\nonumber\\
  &\left.4 C_{\zeta5} \bar{\zeta}^2 \dot{H}
  -4 C_{\zeta6} \bar{\zeta}^2 \dot{H}
  +2
  \dot{D_{\zeta4}} \dot{\bar{\zeta}}+2 D_{\zeta4} \ddot{\bar{\zeta}}+4 D_{\zeta4} \bar{\zeta}
  H^2+2 D_{\zeta4} \dot{\bar{\zeta}} H+8 D_{\zeta7} \bar{\zeta} H^2\right.\nonumber\\
  &\left.-12 D_{\zeta7} \dot{\bar{\zeta}} H+4
  D_{\zeta9} \bar{\zeta} \dot{H}
  -32 H^2 C_1+16 \dot{H}
  C_5\right)\nonumber\\
A_5=&\frac{1}{16 \dot{\bar{\zeta}}}\left(6 B_{\zeta} \dot{H} \bar{\zeta}^3+12 B_{\zeta} H \dot{\bar{\zeta}} \bar{\zeta}^2+12
  C_{\zeta1} H \ddot{\bar{\zeta}} \bar{\zeta}^2+4 \dot{C}_{\zeta1}
  \ddot{\bar{\zeta}} \bar{\zeta}^2+4 C_{\zeta1} \dddot{\bar{\zeta}}
  \bar{\zeta}^2+8 C_{\zeta1} \dot{\bar{\zeta}} \ddot{\bar{\zeta}} \bar{\zeta}\right.\nonumber\\
  &\left.-24 C_{\zeta2}
  H^2 \dot{\bar{\zeta}} \bar{\zeta}^2
  +12 C_{\zeta2} H \dot{H} \bar{\zeta}^3-12 C_{\zeta2}
  \dot{H} \dot{\bar{\zeta}} \bar{\zeta}^2+24 C_{\zeta2} H \dot{\bar{\zeta}}^2 \bar{\zeta} - 24
  C_{\zeta3} H^2 \dot{\bar{\zeta}} \bar{\zeta}^2-24 \dot{C}_{\zeta3} H \dot{\bar{\zeta}}
  \bar{\zeta}^2\right.\nonumber\\
  &\left.+24 C_{\zeta3} \dot{H} \dot{\bar{\zeta}} \bar{\zeta}^2
  -48 C_{\zeta3} H\dot{\bar{\zeta}}^2 \bar{\zeta} - 18 C_{\zeta4} H^2 \dot{\bar{\zeta}} \bar{\zeta}^2-18
  C_{\zeta4} H \dot{H} \bar{\zeta}^3-6 \dot{C}_{\zeta4} \dot{H} \bar{\zeta}^3-6 C_{\zeta4}
  \ddot{H} \bar{\zeta}^3\right.\nonumber\\
  &\left.-6 \dot{C}_{\zeta4} H \dot{\bar{\zeta}} \bar{\zeta}^2
  -24C_{\zeta4} \dot{H} \dot{\bar{\zeta}} \bar{\zeta}^2-
  12 C_{\zeta4} H \dot{\bar{\zeta}}^2
  \bar{\zeta} - 60 C_{\zeta5} H^2 \dot{\bar{\zeta}} \bar{\zeta}^2-12 C_{\zeta5} H \dot{H}
  \bar{\zeta}^3-12 C_{\zeta6} H^2 \dot{\bar{\zeta}} \bar{\zeta}^2\right.\nonumber\\
  &\left.-36 C_{\zeta6} H \dot{H}
  \bar{\zeta}^3
  +42 D_{\zeta4} H^2 \dot{\bar{\zeta}} \bar{\zeta}
  +12 D_{\zeta4} H \dot{H}
  \bar{\zeta}^2+6 D_{\zeta4} \ddot{H} \bar{\zeta}^2+6 \dot{D}_{\zeta4} H
  \dot{\bar{\zeta}} \bar{\zeta}+6 D_{\zeta4} \dot{H} \dot{\bar{\zeta}} \bar{\zeta}\right.\nonumber\\
  &\left.-6 D_{\zeta4}
  H \dot{\bar{\zeta}}^2+24 D_{\zeta7} H \dot{H} \bar{\zeta}^2
  -48 D_{\zeta9} H^2
  \dot{\bar{\zeta}} \bar{\zeta}
  -96 H^2 \dot{\bar{\zeta}} C_5-\dot{\bar{\zeta}}
  \dot{\varphi}^2\right)\nonumber\\
B_2=&D_{\zeta7}(H\bar{\zeta} - \dot{\bar{\zeta}})-4HC_1\nonumber\\
B_2=&\frac{1}{4} \left(-\dot{C_{\zeta5}} \bar{\zeta}^2-2 C_{\zeta5} \dot{\bar{\zeta}} \bar{\zeta}-C_{\zeta5}
  \bar{\zeta}^2 H+\dot{C_{\zeta6}} \bar{\zeta}^2+2 C_{\zeta6} \dot{\bar{\zeta}} \bar{\zeta}+C_{\zeta6} \zeta
  ^2 H+D_{\zeta7} \dot{\bar{\zeta}}-\dot{D}_{\zeta9} \bar{\zeta} - D_{\zeta9} \dot{\bar{\zeta}
  }\right.\nonumber\\
  &\left.-D_{\zeta9} \bar{\zeta} H
  -4 HC_1 -4 HC_5 \right)\nonumber\\
B_3=&\frac{1}{4} \left(B_{\zeta} \bar{\zeta}^2-2 C_{\zeta2} \dot{\bar{\zeta}} \bar{\zeta} + 2 C_{\zeta2} \bar{\zeta}^2 H+4 C_{\zeta3} \bar{\zeta}^2 H-4 C_{\zeta6} \bar{\zeta}^2 H-\dot{D_{\zeta4}} \bar{\zeta} -3
  D_{\zeta4} \bar{\zeta} H+4 D_{\zeta7} \bar{\zeta} H\right.\nonumber\\
  &\left.+16 C_5 H\right)\nonumber\\
B_4=&-4HC_5-\frac{1}{2} \bar{\zeta} \left(-2 C_{\zeta2} \dot{\bar{\zeta}} + 2 C_{\zeta2} \bar{\zeta} H+2
  \dot{C}_{\zeta3} \bar{\zeta} + 4 C_{\zeta3} \dot{\bar{\zeta}} +4 C_{\zeta3} \bar{\zeta} H+2 C_{\zeta5}
  \bar{\zeta} H-2 C_{\zeta6} \bar{\zeta} H\right.\nonumber\\
  &\left.-3 D_{\zeta4} H
  -\dot{D}_{\zeta4}+2 D_{\zeta7} H+2
  D_{\zeta9} H\right)\nonumber\\
 C_1=&\frac{1}{4H} \left(-\dot{C}_{\zeta5} \bar{\zeta}^2-2 C_{\zeta5} \dot{\bar{\zeta}} \bar{\zeta} -C_{\zeta5}
  \bar{\zeta}^2 H+\dot{C_{\zeta6}} \bar{\zeta}^2+2 C_{\zeta6} \dot{\bar{\zeta}} \bar{\zeta} + C_{\zeta6} \bar{\zeta}
  ^2 H+D_{\zeta7} \dot{\bar{\zeta}} - \dot{D}_{\zeta9} \bar{\zeta} -D_{\zeta9} \dot{\bar{\zeta}} \right.\nonumber\\
  &\left.-D_{\zeta9} \bar{\zeta} H-4 \dot{C}_6-4 C_5 H\right)\nonumber\\
C_2=&-C_1\nonumber\\
C_3=&-2C_1\nonumber\\
C_4=&2C_1\nonumber\\
C_7=&\frac{1}{4}\left(-4C_5+C_{\zeta5}\bar{\zeta}^2+C_{\zeta6}\bar{\zeta}^2\right)\nonumber\\
C_7=&\frac{1}{2}\left(4C_5+D_{\zeta7}\bar{\zeta}+D_{\zeta9}\bar{\zeta}+C_{\zeta5}\bar{\zeta}^2-C_{\zeta6}\bar{\zeta}^2\right)\nonumber\\
C_8=&-\frac{1}{2}\left(4C_5+D_{\zeta7}\bar{\zeta}+D_{\zeta9}\bar{\zeta}+C_{\zeta5}\bar{\zeta}^2-C_{\zeta6}\bar{\zeta}^2\right)\nonumber\\
C_9=&\frac{1}{4}\bar{\zeta}\left(C_{\zeta4}\bar{\zeta}-D_{\zeta4}\right)\nonumber\\
C_{10}=&\frac{1}{4}\bar{\zeta}\left(C_{\zeta2}\bar{\zeta}+C_{\zeta3}\bar{\zeta}-D_{\zeta4}\right)\nonumber\\
C_{11}=&-\frac{1}{2}\left(4C_5+2D_{\zeta9}\bar{\zeta}+C_{\zeta5}\bar{\zeta}^2+C_{\zeta6}\bar{\zeta}^2\right)\nonumber\\
C_{12}=&\frac{1}{2}\left(4C_5+2D_{\zeta9}\bar{\zeta}+C_{\zeta5}\bar{\zeta}^2+C_{\zeta6}\bar{\zeta}^2\right)\nonumber\\
C_{13}=&-\left(4C_5+D_{\zeta9}\bar{\zeta}+C_{\zeta5}\bar{\zeta}^2-C_{\zeta6}\bar{\zeta}^2\right)\nonumber\\
C_{14}=&4C_5+D_{\zeta9}\bar{\zeta}+C_{\zeta5}\bar{\zeta}^2-C_{\zeta6}\bar{\zeta}^2\nonumber\\
C_{15}=&C_{\zeta3}\bar{\zeta}^2\nonumber\\
C_{16}=&\frac{1}{2}\left(D_{\zeta4}\bar{\zeta}-2C_{\zeta3}\bar{\zeta}^2\right)\nonumber\\
C_{17}=&\frac{1}{4}C_{\zeta1}\bar{\zeta}^2.
\end{align}
\end{widetext}
For the $A_{\zeta\,n}$, $B_{\zeta\,n}$, $C_{\zeta\,n}$, and $D_{\zeta\,n}$:
\begin{widetext}
\begin{align}
  A_{\zeta1}=&-\frac{1}{2 \bar{\zeta}}\left( -2A_{\zeta3}+2 C_{\zeta2} \ddot{\bar{\zeta}}+6 C_{\zeta2} \bar{\zeta}
  H^2-6 C_{\zeta2} \dot{\bar{\zeta}} H+12 C_{\zeta3} \bar{\zeta} H^2-6 C_{\zeta3} \dot{\bar{\zeta}} H+6
  C_{\zeta3} \bar{\zeta} \dot{H}\right.\nonumber\\
  &\left.-3 C_{\zeta4} \dot{\bar{\zeta}} H-3 C_{\zeta4} \bar{\zeta} \dot{H}-12
  C_{\zeta5} \bar{\zeta} H^2-3 \dot{D}_{\zeta4} H+3 D_{\zeta4} \dot{H}+12 D_{\zeta7} H^2-12
  D_{\zeta9} H^2\right)\nonumber\\
  A_{\zeta2}=&-\frac{1}{2 \bar{\zeta}}\left(-B_{\zeta} \dot{\bar{\zeta}}+2 C_{\zeta2} \bar{\zeta} H^2-2 C_{\zeta2} \dot{\bar{\zeta}} H-2
  \dot{C_{\zeta3}} \dot{\bar{\zeta}}
  -2 C_{\zeta3} \ddot{\bar{\zeta}}+4 C_{\zeta3} \bar{\zeta}
  H^2+2 \dot{C_{\zeta3}} \bar{\zeta} H-\right.\nonumber\\
  &\left.2 C_{\zeta3} \dot{\bar{\zeta}} H+2 C_{\zeta3} \bar{\zeta}
  \dot{H}-C_{\zeta4} \ddot{\bar{\zeta}}+2 C_{\zeta5} \dot{\bar{\zeta}} H-2 C_{\zeta5} \bar{\zeta}
  \dot{H}+6 C_{\zeta6} \dot{\bar{\zeta}} H+6 C_{\zeta6} \bar{\zeta} \dot{H}-4 D_{\zeta9} \dot{H}\right)\nonumber\\
  A_{\zeta3}=&\frac{1}{2\dot{\bar{\zeta}}}\left(3 B_{\zeta} \bar{\zeta}^2 \dot{H}+6 B_{\zeta} \bar{\zeta} \dot{\bar{\zeta}} H+2 \dot{C_{\zeta1}} \zeta
  \ddot{\bar{\zeta}}+2 C_{\zeta1} \dot{\bar{\zeta}} \ddot{\bar{\zeta}}+2 C_{\zeta1} \bar{\zeta}
  \dddot{\bar{\zeta}}+6 C_{\zeta1} \bar{\zeta} \ddot{\bar{\zeta}} H\right.\nonumber\\
  &\left.+6 C_{\zeta2} \bar{\zeta}
  \dot{\bar{\zeta}} H^2+6 C_{\zeta2} \bar{\zeta}^2 \dot{H} H-6 C_{\zeta2} \dot{\bar{\zeta}}^2 H-6
  C_{\zeta2} \bar{\zeta} \dot{\bar{\zeta}} \dot{H}+12 C_{\zeta3} \bar{\zeta} \dot{\bar{\zeta}} H^2-6
  C_{\zeta3} \dot{\bar{\zeta}}^2 H\right.\nonumber\\
  &\left.+6 C_{\zeta3} \bar{\zeta} \dot{\bar{\zeta}} \dot{H}-9 C_{\zeta4}
  \bar{\zeta} \dot{\bar{\zeta}} H^2-9 C_{\zeta4} \bar{\zeta}^2 \dot{H} H-3 \dot{C_{\zeta4}} \bar{\zeta}^2
  \dot{H}-3 C_{\zeta4} \bar{\zeta}^2 \ddot{H}-3 C_{\zeta4} \dot{\bar{\zeta}}^2 H\right.\nonumber\\
  &\left.-3
  \dot{C}_{\zeta4} \bar{\zeta} \dot{\bar{\zeta}} H-9 C_{\zeta4} \bar{\zeta} \dot{\bar{\zeta}} \dot{H}-18
  C_{\zeta5} \bar{\zeta} \dot{\bar{\zeta}} H^2-6 C_{\zeta5} \bar{\zeta}^2 \dot{H} H-18 C_{\zeta6} \zeta
  \dot{\bar{\zeta}} H^2-18 C_{\zeta6} \bar{\zeta}^2 \dot{H} H\right.\nonumber\\
  &\left.+6 D_{\zeta4} \bar{\zeta} \dot{H} H-3
  \dot{D}_{\zeta4} \dot{\bar{\zeta}} H+3 D_{\zeta4} \dot{\bar{\zeta}} \dot{H}+3 D_{\zeta4} \bar{\zeta}
  \ddot{H}+12 D_{\zeta7} \dot{\bar{\zeta}} H^2+12 D_{\zeta7} \bar{\zeta} \dot{H} H\right.\nonumber\\
  &\left.-12
  D_{\zeta9} \dot{\bar{\zeta}} H^2\right)\nonumber\\
  A_{\zeta4}=&\frac{1}{2} \left(-\dot{B_{\zeta}} \bar{\zeta}-B_{\zeta} \dot{\bar{\zeta}}-3 B_{\zeta} \bar{\zeta}
  H+2 \dot{C}_{\zeta2} \dot{\bar{\zeta}}+2 C_{\zeta2} \ddot{\bar{\zeta}}-6 C_{\zeta2} \bar{\zeta}
  H^2-2 \dot{C}_{\zeta2} \bar{\zeta} H+4 C_{\zeta2} \dot{\bar{\zeta}} H\right.\nonumber\\
  &\left.-2 C_{\zeta2} \bar{\zeta}
  \dot{H}+10 C_{\zeta5} \bar{\zeta} H^2+4 \dot{C_{\zeta5}} \bar{\zeta} H+4 C_{\zeta5} \dot{\bar{\zeta}}
  H+2 C_{\zeta5} \bar{\zeta} \dot{H}+6 C_{\zeta6} \bar{\zeta} H^2+6 C_{\zeta6} \bar{\zeta} \dot{H}\right.\nonumber\\
  &\left.+3
  D_{\zeta4} H^2+4 \dot{D}_{\zeta4} H+D_{\zeta4} \dot{H}+\ddot{D_{\zeta4}}-16
  D_{\zeta7} H^2-4 \dot{D}_{\zeta7} H-4 D_{\zeta7} \dot{H}+8 D_{\zeta9} H^2+4
  \dot{D}_{\zeta9} H\right)\nonumber\\
  A_{\zeta5}=&B_{\zeta} \dot{\bar{\zeta}}+2 \dot{C}_{\zeta3} \dot{\bar{\zeta}}+2 C_{\zeta3} \ddot{\bar{\zeta}}-4
  C_{\zeta3} \bar{\zeta} H^2-2 \dot{C}_{\zeta3} \bar{\zeta} H+2 C_{\zeta3} \dot{\bar{\zeta}} H-2
  C_{\zeta3} \bar{\zeta} \dot{H}+C_{\zeta4} \ddot{\bar{\zeta}}\nonumber\\
  &-2 C_{\zeta5} \dot{\bar{\zeta}} H
  +2
  C_{\zeta5} \bar{\zeta} \dot{H}-6 C_{\zeta6} \dot{\bar{\zeta}} H-6 C_{\zeta6} \bar{\zeta}
  \dot{H}-D_{\zeta4} H^2-\dot{D}_{\zeta4} H+4 D_{\zeta7} H^2+4 D_{\zeta9} \dot{H}\nonumber\\
  B_{\zeta1}=&\frac{3}{2}H\left(D_{\zeta4}+2C_{\zeta3}\bar{\zeta}\right)\nonumber\\
  B_{\zeta2}=&\frac{1}{2}\left(HD_{\zeta4}-4HD_{\zeta9}-2\dot{C}_{\zeta3}\bar{\zeta}+B_{\zeta}\bar{\zeta}-2HC_{\zeta3}\bar{\zeta}-4HC_{\zeta5}\bar{\zeta}-4C_{\zeta3}\dot{\bar{\zeta}}\right)\nonumber\\
  B_{\zeta3}=&\frac{1}{2}\left(\dot{D}_{\zeta4}+3HD_{\zeta4}-4HD_{\zeta7}-B_{\zeta}\bar{\zeta}-2HC_{\zeta2}\bar{\zeta}+2HC_{\zeta5}\bar{\zeta}+6HC_{\zeta6}\bar{\zeta}+2C_{\zeta2}\dot{\bar{\zeta}}\right)\nonumber\\
  B_{\zeta4}=&-\dot{D}_{\zeta4}-2HD_{\zeta4}+2HD_{\zeta7}+2HC_{\zeta2}\bar{\zeta}-2C_{\zeta2}\dot{\bar{\zeta}}\nonumber\\
  B_{\zeta5}=&2\left(HD_{\zeta9}+3HC_{\zeta5}\bar{\zeta}+HC_{\zeta6}\bar{\zeta}+C_{\zeta3}\dot{\bar{\zeta}}\right)\nonumber\\
  B_{\zeta6}=&\dot{D}_{\zeta9}-HD_{\zeta7}+2HD_{\zeta9}+\dot{C}_{\zeta5}\bar{\zeta}-\dot{C}_{\zeta6}\bar{\zeta}+2HC_{\zeta5}\bar{\zeta}-2HC_{\zeta6}\bar{\zeta}+C_{\zeta5}\dot{\bar{\zeta}}-C_{\zeta6}\dot{\bar{\zeta}}\nonumber\\
  B_{\zeta7}=&-\dot{D}_{\zeta9}+HD_{\zeta7}-2HD_{\zeta9}-2\dot{C}_{\zeta5}\bar{\zeta}-4HC_{\zeta5}\bar{\zeta}-2C_{\zeta5}\dot{\bar{\zeta}}\nonumber\\
  D_{\zeta1}=&C_{\zeta1}\bar{\zeta}\nonumber\\
  D_{\zeta2}=&\frac{1}{2}\left(-D_{\zeta4}+C_{\zeta4}\bar{\zeta}\right)\nonumber\\
  D_{\zeta3}=&\frac{1}{2}\left(-D_{\zeta4}+C_{\zeta2}\bar{\zeta}\right)\nonumber\\
  D_{\zeta5}=&\frac{1}{2}\bar{\zeta}\left(-2C_{\zeta3}+C_{\zeta4}\right)\nonumber\\
  D_{\zeta6}=&2C_{\zeta3}\bar{\zeta}\nonumber\\
  D_{\zeta8}=&-D_{\zeta7}\nonumber\\
  D_{\zeta10}=&-D_{\zeta9}-C_{\zeta5}\bar{\zeta}+C_{\zeta6}\bar{\zeta}\nonumber\\
  D_{\zeta11}=&D_{\zeta9}+2C_{\zeta5}\bar{\zeta}\nonumber\\
  D_{\zeta12}=&-D_{\zeta9}.
\end{align}
\end{widetext}
In addition, a Friedmann-like equation analogous to (\ref{fried1}) is found:
\begin{align}
  B_{\zeta}=&-\frac{1}{\bar{\zeta}\dot{\bar{\zeta}}}\left(-2 C_{\zeta2} \dot{\bar{\zeta}}^2+2 C_{\zeta2} \dot{\bar{\zeta}} \bar{\zeta} H+C_{\zeta4}
  \ddot{\bar{\zeta}} \bar{\zeta}-2 C_{\zeta5} \bar{\zeta}^2 \dot{H} \right. \nonumber \\ & \left. -2 C_{\zeta5} \dot{\bar{\zeta}}
  \bar{\zeta} H-2 C_{\zeta6} \bar{\zeta}^2 \dot{H}-6 C_{\zeta6} \dot{\bar{\zeta}} \bar{\zeta}
  H-\dot{D}_{\zeta4} \dot{\bar{\zeta}}-D_{\zeta4} \ddot{\bar{\zeta}}\right.\nonumber\\
  &\left.-D_{\zeta4} \dot{\bar{\zeta}}
  H+4 D_{\zeta7} \dot{\bar{\zeta}} H-16 C_6 \dot{H}+\dot{\bar{\varphi}}^2\right).\label{VTfried}
\end{align}

\subsection{Dictionary for scalar perturbations}\label{appendixVT2}
The following dictionary of parameters for the action for scalar perturbations for vector-tensor gravity models, given by eq.~(\ref{VTscalar}), is provided:
\begin{widetext}
\begin{align}
  T_{\Phi^2}=&3H^2\left(3\left(C_{\zeta5}+C_{\zeta6}\right)\bar{\zeta}^2-M^2\right)+3H\left(D_{\zeta4}-C_{\zeta4}\bar{\zeta}\right)\dot{\bar{\zeta}}+C_{\zeta1}\dot{\bar{\zeta}}^2+\frac{1}{2}\dot{\bar{\varphi}}^2\\
  T_{\partial^2\Phi^2}=&C_{\zeta3}\bar{\zeta}^2\\
  T_{\dot{\Psi}^2}=&9\left(C_{\zeta5}+C_{\zeta6}\right)\bar{\zeta}^2-3M^2\\
  T_{\partial^2\Psi^2}=&(1+\alpha_T) M^2\\
  T_{\Phi\Psi}=&36H\frac{\bar{\zeta}}{\dot{\bar{\zeta}}}\dot{\bar{\varphi}}^2\\
  T_{\dot{\Psi}\Phi}=&3\left(2H\left(3\left(C_{\zeta5}+C_{\zeta6}\right)\bar{\zeta}^2-M^2\right)+\left(D_{\zeta4}-C_{\zeta4}\bar{\zeta}\right)\dot{\bar{\zeta}}\right)\\\
  T_{\partial\Phi\partial\Psi}=&-2M^2+4\bar{\zeta}\left(D_{\zeta9}+2C_{\zeta5}\bar{\zeta}\right)\\
  T_{Z_0\Phi}=&\frac{1}{\dot{\bar{\zeta}}}\left(\ddot{\bar{\zeta}} \left(2 C_{\zeta1} \dot{\bar{\zeta}}-3 C_{\zeta4} H \bar{\zeta}+3
  D_{\zeta4} H\right)-3 \dot{H} \left(\dot{\bar{\zeta}} (C_{\zeta4}
  \bar{\zeta}-D_{\zeta4})+2 H \left(M^2-3 \bar{\zeta}^2
  (C_{\zeta5}+C_{\zeta6})\right)\right)-3 H \dot{\bar{\varphi}}^2\right)\\
  T_{\dot{Z}_0\Phi}=&3 H \left(C_{\zeta4}\bar{\zeta}-D_{\zeta4}\right)-2 C_{\zeta1} \dot{\bar{\zeta}}\\
  T_{\partial Z_0\partial\Phi}=&-D_{\zeta4}\\
  T_{Z_0\Psi}=&\frac{3}{\dot{\bar{\zeta}}^2}\left(\dot{\bar{\zeta}} \left(-3 C_{\zeta4} H \bar{\zeta} \ddot{\bar{\zeta}}-\dot{C_{\zeta4}}
  \bar{\zeta} \ddot{\bar{\zeta}}-C_{\zeta4} \bar{\zeta}
  \dot{\ddot{\bar{\zeta}}}+\dot{H} \left(-2 H \left((\alpha_M+3) M^2-9
  \bar{\zeta}^2 (C_{\zeta5}+C_{\zeta6})\right)+6 \dot{C}_{\zeta5} \bar{\zeta}^2+6
  \dot{C_{\zeta6}} \bar{\zeta}^2\right)\right.\right.\nonumber\\
  &\left.+6 C_{\zeta5} \ddot{H} \bar{\zeta}^2+6
  C_{\zeta6} \ddot{H} \bar{\zeta}^2+3 D_{\zeta4} H
  \ddot{\bar{\zeta}}+\dot{D}_{\zeta4} \ddot{\bar{\zeta}}+D_{\zeta4}
  \dot{\ddot{\bar{\zeta}}}-2 \ddot{H} M^2+3 H \dot{\bar{\varphi}}^2\right)+\ddot{\bar{\zeta}} \left(\ddot{\bar{\zeta}} (C_{\zeta4}
  \bar{\zeta}-D_{\zeta4})\right.\nonumber\\
  &\left.\left.+2 \dot{H} \left(M^2-3 \bar{\zeta}^2
  (C_{\zeta5}+C_{\zeta6})\right)+\dot{\bar{\varphi}}^2\right)+\dot{\bar{\zeta}}^2 \left(12
  \dot{H} \bar{\zeta} (C_{\zeta5}+C_{\zeta6})-C_{\zeta4} \ddot{\bar{\zeta}}\right)\right)\\
  T_{\dot{Z}_0\Psi}=&\frac{3}{\dot{\bar{\zeta}}}\left(\ddot{\bar{\zeta}} (C_{\zeta4} \bar{\zeta}-D_{\zeta4})+2 \dot{H} \left(M^2-3
  \bar{\zeta}^2 (C_{\zeta5}+C_{\zeta6})\right)+\dot{\bar{\varphi}}^2)\right)\\
  T_{\dot{Z}_0\dot{\Psi}}=&3\left(C_{\zeta4}\bar{\zeta}-D_{\zeta4}\right)\\
  T_{\partial Z_0\partial\Psi}=&\frac{2}{\dot{\bar{\zeta}}}\left(4 C_{\zeta5} \bar{\zeta} \left(H \bar{\zeta}+2 \dot{\bar{\zeta}}\right)+4 \dot{C_{\zeta5}}
  \bar{\zeta}^2+2 D_{\zeta9} \left(H \bar{\zeta}+\dot{\bar{\zeta}}\right)+2 \dot{D_{\zeta9}}
  \bar{\zeta}+\alpha_M (-H) M^2+\alpha_T H M^2\right)\\
  T_{Z_0^2}=&\frac{1}{2\dot{\bar{\zeta}}^2}\left(\dot{\bar{\zeta}} \left(2 \left(\ddot{\bar{\zeta}} \left(3 C_{\zeta1}
  H+\dot{C}_{\zeta1}\right)+C_{\zeta1} \dddot{\bar{\zeta}}\right)+3
  \ddot{H} (D_{\zeta4}-C_{\zeta4} \bar{\zeta})\right)\right.\nonumber\\
  &\left.-3 \dot{H}
  \left(\dot{\bar{\zeta}} \left(3 C_{\zeta4} H \bar{\zeta}+\dot{C}_{\zeta4}
  \bar{\zeta}-3 D_{\zeta4} H-\dot{D}_{\zeta4}\right)+\ddot{\bar{\zeta}}
  (C_{\zeta4} \bar{\zeta}-D_{\zeta4})+C_{\zeta4} \dot{\bar{\zeta}}^2+\dot{\bar{\varphi}}^2\right)\right.\nonumber\\
  &\left.-6 \dot{H}^2 \left(M^2-3 \bar{\zeta}^2
  (C_{\zeta5}+C_{\zeta6})\right)\right)\\
  T_{\dot{Z}_0^2}=&C_{\zeta1}\\
  T_{\partial^2Z_0^2}=&C_{\zeta2}\\
  T_{\partial\dot{Z_0}\partial Z_1}=&C_{\zeta4}\\
  T_{\partial Z_0\partial Z_1}=&-\frac{1}{\bar{\zeta}\dot{\bar{\zeta}}}\left(-2 C_{\zeta2} \dot{\bar{\zeta}}^2+C_{\zeta4} \bar{\zeta} \ddot{\bar{\zeta}}+8
  C_{\zeta5} H^2 \bar{\zeta}^2+8 \dot{C}_{\zeta5} H \bar{\zeta}^2-6 C_{\zeta5}
  \dot{H} \bar{\zeta}^2+16 C_{\zeta5} H \bar{\zeta} \dot{\bar{\zeta}} -6 C_{\zeta6}
  \dot{H} \bar{\zeta}^2-D_{\zeta4} H \dot{\bar{\zeta}}-\dot{D}_{\zeta4}
  \dot{\bar{\zeta}}\right.\nonumber\\
 &\left.-D_{\zeta4} \ddot{\bar{\zeta}}+4 D_{\zeta9} H^2
  \bar{\zeta}+4 \dot{D}_{\zeta9} H \bar{\zeta}+4 D_{\zeta9} H \dot{\bar{\zeta}}+2 \left(\alpha_T-\alpha_M\right) H^2 M^2+2 \dot{H}M^2+\dot{\bar{\varphi}}^2\right)\\
  T_{\partial\Phi\partial Z_1}=&-2H\left(D_{\zeta9}+\left(C_{\zeta5}-3C_{\zeta6}\right)\bar{\zeta}\right)-\left(2C_{\zeta3}+C_{\zeta4}\right)\dot{\bar{\zeta}}\\
  T_{\partial\Phi\partial\dot{Z}_1}=&2C_{\zeta3}\bar{\zeta}\\
  T_{\dot{\Psi}\dot{Z}_1}=&\frac{2H}{\dot{\bar{\zeta}}}\left(4 C_{\zeta5} \bar{\zeta} \left(H \bar{\zeta}+2 \dot{\bar{\zeta}}\right)+4 \dot{C}_{\zeta5}
  \bar{\zeta}^2+2 D_{\zeta9} \left(H \bar{\zeta}+\dot{\bar{\zeta}}\right)+2 \dot{D}_{\zeta9}
  \bar{\zeta}-\alpha_MH M^2+\alpha_T H M^2\right)\\
  T_{\partial\Psi\partial Z_1}=&2 \left(C_{\zeta5} H \bar{\zeta}+\dot{C}_{\zeta5} \bar{\zeta}+C_{\zeta5} \dot{\bar{\zeta}}-3
  C_{\zeta6} H \bar{\zeta}-3 \dot{C_{\zeta6}} \bar{\zeta}-3 C_{\zeta6} \dot{\bar{\zeta}}+2
  D_{\zeta9} H+2 \dot{D}_{\zeta9}\right)\\
  T_{\partial\dot{\Psi}\partial Z_1}=&4 D_{\zeta9}+(2 C_{\zeta5}-6 C_{\zeta6})\bar{\zeta}\\
  T_{\partial^2 Z_1^2}=&\frac{1}{2\bar{\zeta}^2}\left(2 C_{\zeta1} \dot{\bar{\zeta}}^2+2 C_{\zeta3} H \bar{\zeta} \dot{\bar{\zeta}}+2
  \dot{C_{\zeta3}} \bar{\zeta} \dot{\bar{\zeta}}+2 C_{\zeta3} \bar{\zeta}
  \ddot{\bar{\zeta}}-8 C_{\zeta5} H^2 \bar{\zeta}^2-8 \dot{C}_{\zeta5} H
  \bar{\zeta}^2\right.\nonumber\\
  &+8 C_{\zeta5} \dot{H} \bar{\zeta}^2-16 C_{\zeta5} H \bar{\zeta}
  \dot{\bar{\zeta}}+D_{\zeta4} H \dot{\bar{\zeta}}+\dot{D}_{\zeta4}
  \dot{\bar{\zeta}}+D_{\zeta4} \ddot{\bar{\zeta}}-4 D_{\zeta9} H^2
  \bar{\zeta}-4 \dot{D}_{\zeta9} H \bar{\zeta}\nonumber\\
  &\left.-4 D_{\zeta9} H \dot{\bar{\zeta}}+4
  D_{\zeta9} \dot{H} \bar{\zeta}+2 \left(\alpha_M-\alpha_T\right) H^2 M^2-2 \dot{H} M^2-\dot{\bar{\varphi}}^2\right)\\
  T_{\partial^2 \dot{Z}_1^2}=&C_{\zeta3}\\
  T_{\partial^4 Z_1^2}=&C_{\zeta5}+C_{\zeta6}.
\end{align}
\end{widetext}
We see that all these functions $T$ depend only on 9 combinations of the 10 free parameters:
\begin{align}
&M^2,\;C_{\zeta1},\;C_{\zeta2},\;C_{\zeta3},\;C_{\zeta4},\;C_{\zeta5}+C_{\zeta6},\;D_{\zeta4},\;\alpha_T,\nonumber\\
&D_{\zeta9}+2\bar{\zeta}C_{\zeta5},
\end{align}
and hence only these combinations can be constrained by observing the effect of scalar perturbations in the Universe.

\subsection{Dictionary for vector perturbations}\label{appendixVT3}
The following dictionary of parameters for the action for scalar perturbations in vector-tensor gravity models, given by eq.~(\ref{VTactionVector}), is provided:
\begin{widetext}
\begin{align}
T_{\partial^2 N^2}=&\frac{1}{4}M^2\\
  T_{\partial N \partial \delta Z}=&-D_{\zeta9}-2\bar{\zeta}C_{\zeta5}\\
  T_{\delta Z^2}=&\frac{1}{2\bar{\zeta}^2}\left(2 C_{\zeta2} \dot{\bar{\zeta}}^2-2 C_{\zeta3} H^2 \bar{\zeta}^2-2 C_{\zeta3} \dot{H} \bar{\zeta}^2+2 C_{\zeta3} H \bar{\zeta} \dot{\bar{\zeta}}+2 \dot{C_{\zeta3}} \bar{\zeta}
  \left(\dot{\bar{\zeta}}-H \bar{\zeta}\right)+2 C_{\zeta3} \bar{\zeta}\ddot{\bar{\zeta}}-8 C_{\zeta5} H^2 \bar{\zeta}^2-8 \dot{C}_{\zeta5} H \bar{\zeta}^2\right.\nonumber\\
  &+8 C_{\zeta5}
  \dot{H} \bar{\zeta}^2-16 C_{\zeta5} H \bar{\zeta} \dot{\bar{\zeta}}+D_{\zeta4} H \dot{\bar{\zeta}}+\dot{D}_{\zeta4} \dot{\bar{\zeta}}+D_{\zeta4}\ddot{\bar{\zeta}}-4
  D_{\zeta9} H^2 \bar{\zeta}-4 \dot{D}_{\zeta9} H \bar{\zeta}-4 D_{\zeta9} H \dot{\bar{\zeta}}+4 D_{\zeta9} \dot{H} \bar{\zeta}\nonumber\\
  &\left.+2 \left(\alpha_M-\alpha_T\right) H^2 M^2-2 \dot{H} M^2-\dot{\bar{\varphi}}^2\right)\\
  T_{\delta\dot{Z}^2}=&C_{\zeta3}\\
  T_{\partial^2\delta Z^2}=&C_{\zeta5}.
\end{align}
\end{widetext}

We see that this action depends only on 5 combinations of the free parameters:
\begin{align}
 &M^2, \; C_{\zeta3}, \; C_{\zeta5}, \; D_{\zeta9},\nonumber\\
 &2C_{\zeta2}\dot{\bar{\zeta}}^2+D_{\zeta4} H \dot{\bar{\zeta}}+\dot{D}_{\zeta4} \dot{\bar{\zeta}}+D_{\zeta4}\ddot{\bar{\zeta}}-2\alpha_T H^2M^2,
 \end{align}
and hence only these combinations can be constrained by searching observational signatures of vector perturbations in these models.

\section{Noether Constraints for an axisymmetric Bianchi-I vacuum universe}\label{appendixBianchi}
\subsection{Most general quadratic action}
The tensors used in action (\ref{Sbianchi}), where we again only include tensor terms which will give independent terms in the action, are given by:
\begin{widetext}
\begin{align}
\mathcal{D}^{\mu\nu\alpha\beta}= &D_1 \gamma^{\mu\nu} \gamma^{\alpha\beta}+\gamma^{\mu\nu} \left(D_4 u^\alpha x^\beta+D_2
  u^\alpha u^\beta+D_3 x^\alpha x^\beta\right)+\gamma^{\mu\alpha} \left(D_5 \gamma^{\nu\beta}+D_8
  u^\nu x^\beta+D_6 u^\nu u^\beta+D_7 x^\nu x^\beta\right)+u^\mu D_{14} x^\nu
  u^\alpha x^\beta\nonumber\\
  &+u^\mu u^\nu \left(D_{11} u^\alpha x^\beta+D_9 u^\alpha u^\beta+D_{10}
  x^\alpha x^\beta\right)+x^\mu x^\nu \left(D_{13} x^\alpha u^\beta+D_{12} x^\alpha x^\beta\right)\\
\mathcal{E}^{\mu\nu\alpha\beta\delta}=&\gamma^{\mu\nu}\left(u^\alpha x^\beta u^\delta E_7+x^\alpha x^\beta u^\delta E_9+u^\alpha
  u^\beta x^\delta E_{10}+u^\alpha x^\beta x^\delta E_8+u^\alpha u^\beta u^\delta
  E_5+x^\alpha x^\beta x^\delta E_6\right)+\gamma^{\mu\alpha} \left(u^\nu x^\beta u^\delta
  E_{17}+u^\nu x^\beta x^\delta E_{18}\right)\nonumber\\
  &+\gamma^{\mu\delta} \left(x^\nu u^\alpha u^\beta
  E_{16}+u^\nu u^\alpha x^\beta E_{13}+x^\nu u^\alpha x^\beta E_{14}+u^\nu
  x^\alpha x^\beta E_{15}+u^\nu u^\alpha u^\beta E_{11}+x^\nu x^\alpha x^\beta
  E_{12}\right)+\gamma^{\mu\delta} \gamma^{\nu\alpha} \left(u^\beta E_3+x^\beta E_4\right)\nonumber\\
  &+\gamma^{\mu\nu}
 \gamma^{\alpha\delta} \left(u^\beta E_1+x^\beta E_2\right)+x^\mu x^\nu x^\alpha u^\beta u^\delta
  E_{23}+x^\mu x^\nu x^\alpha u^\beta x^\delta E_{21}+u^\mu u^\nu u^\alpha x^\beta u^\delta
  E_{22}+u^\mu u^\nu u^\alpha x^\beta x^\delta E_{24}\nonumber\\
  &+x^\mu x^\nu u^\alpha u^\beta \left(u^\delta
  E_{20}+x^\delta E_{19}\right)\\
\mathcal{F}^{\mu\nu\alpha\beta\kappa\delta}=&F_1\gamma^{\mu\nu}\gamma^{\alpha\beta}\gamma^{\kappa\delta}+ F_2\gamma^{\mu\alpha}\gamma^{\nu\beta}\gamma^{\kappa\delta}+ F_3\gamma^{\mu\nu}\gamma^{\alpha\kappa}\gamma^{\beta\delta}+ F_4\gamma^{\mu\kappa}\gamma^{\alpha\beta}\gamma^{\nu\delta}
 +\left(F_5 \gamma^{\mu\nu}\gamma^{\alpha\beta}+ F_6\gamma^{\mu\alpha}\gamma^{\nu\beta}\right)u^\kappa u^\delta \nonumber\\
 &+ \left(F_7 \gamma^{\mu\nu}\gamma^{\kappa\delta}+ F_8\gamma^{\mu\kappa}\gamma^{\nu\delta}\right)u^\alpha u^\beta + F_{9}\gamma^{\alpha\beta} u^\mu u^\nu u^\kappa u^\delta+F_{10}\gamma^{\kappa\delta} u^\alpha u^\beta u^\mu u^\nu
+ \left(F_{11}\gamma^{\kappa\delta}\gamma^{\beta \nu}+ F_{12}\gamma^{\kappa\beta}\gamma^{\delta \nu}\right)u^\mu u^\alpha\nonumber\\
&+ \left(F_{13}\gamma^{\alpha\beta}\gamma^{\nu \delta}+ F_{14}\gamma^{\alpha\nu}\gamma^{\delta \beta}\right)u^\mu u^\kappa
+F_{15}\gamma^{\mu\alpha}u^\nu u^\beta u^\kappa u^\delta + F_{16}\gamma^{\mu\kappa}u^\nu u^\beta u^\alpha u^\delta+ F_{17} u^\mu u^\alpha u^\nu u^\beta u^\kappa u^\delta\nonumber\\
&+\left(F_{18} \gamma^{\mu\nu}\gamma^{\alpha\beta}+ F_19\gamma^{\mu\alpha}\gamma^{\nu\beta}\right)x^\kappa x^\delta + \left(F_{20} \gamma^{\mu\nu}\gamma^{\kappa\delta}+ F_{21}\gamma^{\mu\kappa}\gamma^{\nu\delta}\right)x^\alpha x^\beta + F_{22}\gamma^{\alpha\beta} x^\mu x^\nu x^\kappa x^\delta+F_{23}\gamma^{\kappa\delta} x^\alpha x^\beta x^\mu x^\nu
\nonumber\\
&+ \left(F_{24}\gamma^{\kappa\delta}\gamma^{\beta \nu}+ F_{25}\gamma^{\kappa\beta}\gamma^{\delta \nu}\right)x^\mu x^\alpha+ \left(F_{26}\gamma^{\alpha\beta}\gamma^{\nu \delta}+ F_{27}\gamma^{\alpha\nu}\gamma^{\delta \beta}\right)x^\mu x^\kappa
+F_{28}\gamma^{\mu\alpha}x^\nu x^\beta x^\kappa x^\delta + F_{29}\gamma^{\mu\kappa}x^\nu x^\beta x^\alpha x^\delta\nonumber\\
&+ F_{30} x^\mu x^\alpha x^\nu x^\beta x^\kappa x^\delta+\gamma^{\mu\nu}\left(F_{31}\gamma^{\alpha\beta}x^\kappa u^\delta +F_{32}\gamma^{\alpha\kappa}u^\beta x^\delta+F_{33}\gamma^{\alpha\kappa}u^\delta x^\beta+F_{34}\gamma^{\kappa\delta}x^\alpha u^\beta\right)\nonumber\\
&+\gamma^{\mu\alpha}\left(F_{35}\gamma^{\nu\beta}u^\kappa x^\delta +F_{36}\gamma^{\kappa\delta}x^\nu u^\beta\right)+\gamma^{\mu\kappa}\left(F_{37}\gamma^{\alpha\delta}x^\nu u^\beta+F_{38}\gamma^{\nu\delta}x^\alpha u^\beta +F_{39}\gamma^{\nu\alpha}x^\beta u^\delta +F_{40}\gamma^{\nu\alpha}u^\beta x^\delta\right)\nonumber\\
&+ \gamma^{\mu\nu} \left(u^\alpha x^\beta u^\kappa u^\delta F_{55}+u^\alpha u^\beta u^\kappa x^\delta
  F_{54}\right)+\gamma^{\mu\nu} \left(x^\alpha x^\beta u^\kappa u^\delta F_{41}+x^\alpha u^\beta
  x^\kappa u^\delta F_{43}+u^\alpha u^\beta x^\kappa x^\delta F_{42}\right)\nonumber\\
  &+\gamma^{\mu\nu}
  \left(x^\alpha x^\beta x^\kappa u^\delta F_{62}+x^\alpha u^\beta x^\kappa x^\delta
  F_{63}\right)+\gamma^{\mu\alpha} \left(u^\nu x^\beta u^\kappa u^\delta F_{60}+u^\nu u^\beta
  u^\kappa x^\delta F_{59}\right)\nonumber\\
  &+\gamma^{\mu\alpha} \left(x^\nu x^\beta u^\kappa u^\delta
  F_{50}+u^\nu x^\beta u^\kappa x^\delta F_{52}+u^\nu u^\beta x^\kappa
  x^\delta F_{51}\right)+\gamma^{\mu\alpha} \left(x^\nu x^\beta x^\kappa u^\delta F_{67}+x^\nu
  u^\beta x^\kappa x^\delta F_{68}\right)\nonumber\\
  &+\gamma^{\mu\kappa} \left(x^\nu u^\alpha u^\beta u^\delta
  F_{58}+u^\nu u^\alpha x^\beta u^\delta F_{56}+u^\nu u^\alpha u^\beta
  x^\delta F_{57}\right)\nonumber\\
  &+\gamma^{\mu\kappa} \left(x^\nu u^\alpha x^\beta u^\delta F_{49}+u^\nu
  x^\alpha x^\beta u^\delta F_{46}+u^\nu u^\alpha x^\beta x^\delta F_{48}+x^\nu
  u^\alpha u^\beta x^\delta F_{47}\right)\nonumber\\
  &+\gamma^{\mu\kappa} \left(x^\nu x^\alpha x^\beta u^\delta
  F_{65}+u^\nu x^\alpha x^\beta x^\delta F_{66}+x^\nu x^\alpha u^\beta
  x^\delta F_{64}\right)+x^\mu x^\nu x^\alpha u^\beta F_{61} \gamma^{\kappa\delta}+u^\mu
  u^\nu u^\alpha x^\beta F_{53} \gamma^{\kappa\delta}\nonumber\\
  &+\gamma^{\kappa\delta} \left(x^\mu u^\nu
  x^\alpha u^\beta F_{45}+u^\mu u^\nu x^\alpha x^\beta F_{44}\right)+x^\mu u^\nu u^\alpha u^\beta u^\kappa u^\delta F_{70}+x^\mu x^\nu u^\alpha u^\beta u^\kappa
  u^\delta F_{75}+x^\mu u^\nu x^\alpha u^\beta u^\kappa u^\delta F_{76}\nonumber\\
  &+x^\mu x^\nu
  x^\alpha x^\beta u^\kappa u^\delta F_{77}+x^\mu u^\nu u^\alpha u^\beta x^\kappa u^\delta
  F_{74}+x^\mu x^\nu x^\alpha x^\beta x^\kappa u^\delta F_{71}+u^\mu x^\nu x^\alpha
  x^\beta u^\kappa u^\delta F_{81}+u^\mu u^\nu u^\alpha u^\beta x^\kappa u^\delta
  F_{69}\nonumber\\
  &+u^\mu x^\nu u^\alpha x^\beta u^\kappa x^\delta F_{83}+u^\mu u^\nu x^\alpha
  x^\beta u^\kappa x^\delta F_{82}+u^\mu x^\nu x^\alpha x^\beta u^\kappa x^\delta
  F_{78}+u^\mu u^\nu u^\alpha u^\beta x^\kappa x^\delta F_{73}+u^\mu u^\nu u^\alpha
  x^\beta x^\kappa x^\delta F_{84}\nonumber\\
  &+u^\mu x^\nu u^\alpha x^\beta x^\kappa x^\delta
  F_{80}+u^\mu u^\nu x^\alpha x^\beta x^\kappa x^\delta F_{79}+u^\mu x^\nu x^\alpha
  x^\beta x^\kappa x^\delta F_{72},
\end{align}
\end{widetext}
where each of the $D_n$, $E_n$, and $F_n$ are free functions of time.
\subsection{Covariant quantities}
For an anisotropic background, the background space-time will no longer be flat. Thus we need expressions for the Christoffel symbols and curvature tensors of the background \textit{in terms of the background quantities} to properly evaluate the Noether constraints arising from the variation of (\ref{Sbianchi}). The relevant expressions can be shown to be:
\begin{align}
\bar{\nabla}_\mu u_\nu=&H_1 x_\mu x_\nu +H_2\gamma_{\mu\nu}\\
\bar{\nabla}_\mu x_\nu=&H_1 x_\mu u_\nu \\
\bar{\nabla}_\mu\gamma_{\alpha\beta}=&u_\alpha\bar{\nabla}_\mu u_\beta + u_\beta\bar{\nabla}_\mu u_\alpha - x_\alpha\bar{\nabla}x_\beta - x_\beta\bar{\nabla}x_\alpha\\
\bar{R}^\rho_{\,\sigma\mu\nu}=&\dot{H}_1(-u_\mu u^\rho x_{\nu}x_{\sigma}+u_\nu u^\rho x_{\mu}x_{\sigma}+x^\rho x_\nu u_\mu u_\sigma\nonumber\\
&-x^\rho x_\mu u_\nu u_\sigma)+H_1^2(-u_\mu u^\rho x_{\nu}x_{\sigma}+u_\nu u^\rho x_{\mu}x_{\sigma}\nonumber\\
&+x^\rho x_\nu u_\mu u_\sigma-x^\rho x_\mu u_\nu u_\sigma)+\dot{H}_2(-\gamma^\rho_\mu u_\sigma u_\nu\nonumber\\
&+\gamma^\rho_\nu u_\sigma u_\mu + u^\rho u_\nu \gamma_{\mu\sigma} - u^\rho u_\mu \gamma_{\sigma\nu})\nonumber\\
&+H_2^2(-\gamma^\rho_\mu u_\sigma u_\nu+\gamma^\rho_\nu u_\sigma u_\mu + u^\rho u_\nu \gamma_{\mu\sigma} \nonumber\\
&- u^\rho u_\mu \gamma_{\sigma\nu})+H_1H_2(\gamma^\rho_\mu x_\sigma x_\nu -\gamma^\rho_\nu x_\sigma x_\mu\nonumber\\
& -x^\rho x_\nu \gamma_{\mu\sigma}+ x^\rho x_\mu \gamma_{\sigma\nu})\nonumber\\
\bar{R}_{\mu\nu}=&-\left(\dot{H}_1+H_1^2+2\dot{H}_2+2H_2^2\right)u_\mu u_\nu\nonumber\\
&+\left(\dot{H}_1+H_1^2+2H_1H_2\right)x_\mu x_\nu\nonumber\\
& +\left(2H_2^2+H_1H_2+\dot{H}_2\right)\gamma_{\mu\nu}\\
\bar{R}=&2H_1^2+6H_2^2+4H_1H_2+2\dot{H}_1+4\dot{H}_2.
\end{align}
$H_1=\frac{d\log{a}}{dt}$ and $H_2=\frac{d\log{b}}{dt}$ are the two Hubble parameters, $\bar{R}^\rho_{\,\sigma\mu\nu}$ is the Riemann curvature tensor, $\bar{R}_{\mu\nu}=\bar{R}^\rho_{\mu\rho\nu}$ is the Ricci tensor, and $\bar{R}=\bar{g}^{\mu\nu}\bar{R}_{\mu\nu}$ is the Ricci scalar.

\subsection{Solutions}
The following Noether constraints are obtained in Section \ref{Symmetries} for the $D_n$, $E_n$, and $F_n$:
\begin{widetext}
\begin{align}
-F_2=&-\frac{1}{2}F_3=\frac{1}{2}F_4=F_1\nonumber\\
2F_5=&-2F_6=F_7=-F_8=-F_{11}=F_{12}=-\frac{1}{2}F_{13}=\frac{1}{2}F_{14}=F_{41}\nonumber\\
F_9=&F_{10}=F_{15}=F_{16}=F_{17}=0\nonumber\\
-2F_{18}=&2F_{19}=-F_{20}=F_{21}=F_{24}=-F_{25}=\frac{1}{2}F_{26}=-\frac{1}{2}F_{27}=F_{41}\nonumber\\
F_{22}=&F_{23}=F_{28}=F_{29}=F_{30}=0\nonumber\\
F_{31}=&...=F_{40}=0\nonumber\\
F_{42}=&-\frac{1}{2}F_{43}=F_{44}=-F_{45}=-\frac{1}{2}F_{46}=-\frac{1}{2}F_{47}=\frac{1}{2}F_{48}=\frac{1}{2}F_{49}=-F_{50}=-F_{51}=\frac{1}{2}F_{52}=F_{41}\nonumber\\
F_{53}=&...=F_{84}=0\nonumber\\
\frac{1}{2}E_1=&-\frac{1}{4}E_3=F_1H_2\nonumber\\
E_2=&E_4=E_6=E_7=E_{10}=E_{12}=E_{13}=E_{16}=E_{17}=E_{19}=E_{21}=E_{22}=E_{23}=0\nonumber\\
E_5=&E_9=-F_{41}\left(H_1+H_2\right)\nonumber\\
\frac{1}{4}E_8=&E_{11}=-\frac{1}{2}E_{14}=E_{15}=\frac{1}{4}E_{18}=\frac{1}{2}E_{20}=\frac{1}{4}E_{24}=F_{41}H_2\nonumber\\
\frac{1}{2}D_1=&-\frac{1}{4}D_5=F_1H_2^2\nonumber\\
D_2=&-2F_{41}H_2\left(H_1+H_2\right)\nonumber\\
D_3=&2F_{41}H_1H_2\nonumber\\
D_4=&D_8=D_{11}=D_{12}=D_{13}=0\nonumber\\
D_6=&2H_2\left(F_{41}H_1+F_{41}H_2-F_1H_2\right)\nonumber\\
D_7=&2F_{41}H_2\left(H_1-H_2\right)\nonumber\\
2D_9=&D_{10}=4F_{41}H_2\left(H_1+\frac{1}{2}H_2\right)\nonumber\\
D_{14}=&4F_{41}H_2\left(H_1-\frac{1}{2}H_2\right).
\end{align}
\end{widetext}
The following evolution equations are also found:
\begin{align}
\dot{F}_{41}=&0\label{F41dot}\\
\dot{H}_1=&-\left(H_1-H_2\right)\left(H_1+H_2\right)\label{H1dot}\\
\dot{H}_2=&\left(H_1-H_2\right)H_2.\label{H2dot}
\end{align}
eq.~(\ref{F41dot}) requires that $F_{41}$ be a constant, which we relabel as $c_1$. We also relabel $F_{1}=\frac{1}{8}M^2$. Eqs (\ref{H1dot}) and (\ref{H2dot}) can be combined to find:
\begin{align}
H_1=\frac{c_2}{H_2}-\frac{1}{2}H_2,
\end{align}
where $c_2$ is a constant of integration.

\bibliographystyle{apsrev4-1}
\bibliography{cosmology}

\end{document}